\documentclass[aps,pre,amsmath,amssymb]{revtex4-1}
\usepackage{graphicx}
\usepackage{dcolumn}
\usepackage{bm}
\usepackage{hyperref}
\newcommand{\sgn}{\operatorname{sgn}}

\begin{document}
\title{Fluctuating Phases and Fluctuating Relaxation Times in Glass
  Forming Liquids}  
\author{Gcina~A.~Mavimbela} 
\address{Department of Physics, University of Swaziland, P/Bag 4 
  Kwaluseni, Swaziland}
\author{Azita Parsaeian} 
\address{Materials Research Center, Northwestern University, Evanston,
  IL 60208-3108, USA} 
\author{Horacio~E.~Castillo}
\address{Department of Physics and Astronomy, Ohio University, Athens,
  Ohio, USA, 45701} 
\date{\today}

\begin{abstract}
The presence of fluctuating local relaxation times, $\tau_{\vec{r}}(t)$
has been used for some time as a conceptual tool to describe dynamical
heterogeneities in glass-forming systems. However, until now no 
general method is known to extract the full space and time 
dependent $\tau_{\vec{r}}(t)$ from experimental or numerical 
data. Here we report on a new
method for determining the local phase field,
$\phi_{\vec{r}}(t)\equiv\int^{t}\frac{dt'}{\tau_{\vec{r}}(t')}$ from
snapshots $\{\vec{r}(t_i)\}_{i=1...M}$ of the positions of the
particles in a system, and we apply it to extract $\phi_{\vec{r}}(t)$
and $\tau_{\vec{r}}(t)$ from numerical simulations. By
studying how the phase field depends on the number of snapshots, we
find that it is a well defined quantity. By studying fluctuations of
the phase field, we find that they describe heterogeneities well at
long distance scales. 
\end{abstract}

\pacs{64.70.Q-, 61.20.Lc, 61.43.Fs}
%
%
%


\maketitle


\section{Introduction}
As glass-forming liquids enter the supercooled state, relaxation
processes gradually but dramatically slow down, and at the same time
they become non-exponential~\cite{Debenedetti2001, 
  Sillescu1999, Ediger2000}. 
One possible way of understanding this behavior would be to think of
the relaxation function $F(t)$ as the sum of a large number of
exponentially decaying contributions, each with different relaxation
times; with each contribution corresponding to a
particular region in the system~\cite{Sillescu1999, Ediger2000}. This
is one aspect of the picture of {\em dynamical heterogeneity\/}, in
which mesoscopic regions relax differently from each other and from
the bulk~\cite{Debenedetti2001, Sillescu1999, Ediger2000, Russell2000, 
  Weeks2000, Weeks2002, Courtland2003, Cipelletti2003, Keys2007,
  Kob1997, Lacevic2003, Toninelli2005}. In this
framework, it is assumed that the differences between regions are not
frozen in but dynamical, {\em i.e.\/} as the system evolves, the local
relaxation times $\tau_{\vec{r}}(t)$ evolve, so that ``fast'' regions
become ``slow'', and viceversa. This same picture has also been
invoked to explain other phenomena occuring in the same regime, such
as the breakdown of the Stokes-Einstein relation between the diffusion
coefficient and the shear viscosity. Direct evidence for dynamical
heterogeneity has been found in particle tracking experiments in
glassy colloidal systems~\cite{Weeks2000, Weeks2002, Courtland2003}
and granular systems~\cite{Keys2007}, and in numerical
simulations~\cite{Kob1997, Lacevic2003}.

The origin of dynamical heterogeneity is still uncertain, although
some explanations have been proposed~\cite{Toninelli2005}. One of the
proposed explanations is that dynamical heterogeneity emerges as a
consequence of dynamical constraints, which produce non trivial
structure in the space of (space-time) particle
trajectories~\cite{Garrahan2002, Chandler2010}. Another proposal is
that as a liquid is supercooled it eventually undergoes a Random First
Order Transition in which the liquid freezes into a mosaic of
aperiodic crystals~\cite{Xia2001, Lubchenko2007}, which give rise to
the heterogeneity that is observed. A third approach postulates that
dynamical heterogeneity emerges from the presence of soft (Goldstone)
modes associated with a broken continuous symmetry under time
reparametrizations $t \to \phi(t)$~\cite{Chamon2002, Castillo2002,
  Castillo2003, Castillo2007, Castillo2008, Parsaeian2008a,
  Parsaeian2008, Parsaeian2009, Mavimbela2011, Avila2011,
  triangular_long}. 

A quantitative description of dynamical heterogeneity in terms of the
presence of locally fluctuating relaxation times $\tau_{\vec{r}}(t)$ has
in principle some strong advantages. One of them is its simplicity and
intuitive appeal. Another one is that the basic quantities that appear
in this description are intrinsically instantaneous, as opposed to
other common descriptions for which the basic quantities describe
changes in the system over a finite time interval, which complicates
their interpretation. Despite those advantages, however, such a
description has proved elusive. For example, even the question of
experimentally determining the relationship between the lifetime
$\tau_{\rm ex}$ of regions of heterogeneous dynamics and the bulk
$\alpha$-relaxation time $\tau_{\alpha}$ has proved controversial,
with some results indicating that $\tau_{\rm ex}/\tau_{\alpha} \sim
1$, and other results indicating that $\tau_{\rm ex}/\tau_{\alpha} \gg
1$~\cite{Ediger2000}.

In the present work, our goal is twofold. On one hand, we establish a
connection between the presence of time reparametrization fluctuations
and the presence of fluctuating local relaxation times
$\tau_{\vec{r}}(t)$. On the other hand, we introduce a new method to
extract the actual values of those relaxation times
$\tau_{\vec{r}}(t)$ from experimental or numerical data, and test this
method on data from numerical simulations.

Our paper is organized as follows. In Sec.~\ref{sec:equivalence} we
discuss the connection between the presence of time reparametrization
fluctuations and a description of the system in terms of fluctuating
local relaxation times. In Sec.~\ref{sec:method} we explain our
proposed method to extract local relaxation times from numerical or
experimental data. In Sec.~\ref{sec:tests} we discuss the details of
the numerical simulations that we have used to test the method, and we
present the results of those tests. Finally, in
Sec.~\ref{sec:conclusion}, we summarize our results and present some
remarks and ideas for further work.

\section{Connection between fluctuating local relaxation times and
  time reparametrization fluctuations}
\label{sec:equivalence}
At temperatures close to the calorimetric glass transition temperature
$T_g$, the dynamics of glass forming liquids is dramatically different
from the high temperature dynamics~\cite{Ediger2000,Binder-Kob-2005}.
At higher temperatures the relaxation is exponential, a direct
consequence of Brownian-like motion.  By contrast, at lower
temperatures the relaxation is exponential only at the early times
(with fast dynamics) and is followed by a plateau at intermediate
times. This plateau is attributed to the caging effect that results
from having each particle temporarily trapped by its neighbors. At
longer timescales, as the cages start to break and the particles move
distances of the order of the inter-particle distances, the relaxation
function starts to decay. Unlike what happens at high temperatures,
the final decay, {\em i.e.\/} the long time slow dynamics, is usually
not exponential and in many cases it is described well by the
Kohlrausch-Williams-Watts (KWW) ``stretched exponential'' function
\begin{equation}
\label{eq:sexp}
C(t-t') = A \exp\left\{ -[(t-t')/\tau]^{\beta} \right\},
\end{equation}
where we are using some correlation function $C(t-t')$ between 
the states of the system at times $t$ and $t'$ as an example of a
relaxation function.  
Here $A$ and $\beta$ are constants and $\beta$ is sometimes called the "stretching 
exponent" or the "Kohlrausch exponent". Two extreme scenarios were 
proposed to explain the non-exponential behavior~\cite{Ediger2000,Binder-Kob-2005}. 
In one of them the liquid is heterogeneous, the relaxation is exponential in 
each small region, there is 
a spatial distribution of relaxation times $P(\tau)$, and the relaxation 
function is given by
\begin{equation}
\label{eq:het}
C(t-t')=\int_0^{\infty} d\tau P(\tau)\exp\left(-\frac{t-t'}{\tau}\right)
       =\int \frac{d^{d}r}{V}C_{\vec{r}}(t,t')
       \quad\mbox{ with }C_{\vec{r}}(t,t') = C_0
       \exp\left(-\frac{t-t'}{\tau_r}\right), 
\end{equation} 
where $C_{\vec{r}}(t,t')$ is the local two-time correlation 
in a small region around point $\vec{r}$ and $\tau_r$ 
is the corresponding relaxation time.
In the other extreme scenario, the dynamics of the liquid is
homogeneous, the distribution of the relaxation times is a delta function,
and the relaxation in all regions is described by a unique 
function $g(x)$, which is non-exponential:
\begin{equation}
\label{eq:nothet}
C(t-t')=\int^{\infty}_{0}d\tau\delta(\tau-\tau_0)g\left(\frac{t-t'}{\tau}\right)
       =\int \frac{d^{d}r}{V}C_{\vec{r}}(t,t')\quad
       \text{ with }C_{\vec{r}}(t,t')=g\left(\frac{t-t'}{\tau_0}\right).
\end{equation}
Eqs.~(\ref{eq:het}) and (\ref{eq:nothet}) 
can be generalized to allow for the simultaneous presence of both 
heterogeneous relaxation times and non-exponential local relaxation 
functions~\cite{Bohmer-1998}:
\begin{equation}
\label{eq:bohmer}
C(t-t')=\int^{\infty}_{0}d\tau P_g(\tau)g\left(\frac{t-t'}{\tau}\right)
       =\int \frac{d^{d}r}{V}C_{\vec{r}}(t,t')\quad \text{ with } 
       C_{\vec{r}}(t,t')=g\left(\frac{t-t'}{\tau_{\vec{r}}}\right),
\end{equation}
where $P_g(\tau)$ is the probability density of relaxation times compatible 
with the local relaxation function $g(x)$.

If we allow the time difference $t-t'$ to be long enough, we cannot
assume that each local region is characterized by a single relaxation
time $\tau_{\vec{r}}$, because $\tau_{\vec{r}}$ could in principle
have fluctuated over this time interval.  To take this effect into
account, let us divide the time interval $[t',t)$ into $n-1$
  sub-intervals $\{[t_{i-1},t_i)\}_{i=2,...,n}$ such that for each
    sub-interval the fluctuations in $\tau_{\vec{r}}(t)$ are
    negligible.  Hence, for long time intervals, the local correlation
    is given by
\begin{equation}
C_{\vec{r}}(t,t')=g\left[\sum_{i=2}^{n}
                   \frac{t_{i}-t_{i-1}}{\tau_{\vec{r}}(t_{i-1})}\right],
\end{equation}
where  $t_n=t$ and $t_{1}=t'$. In the limit of infinitesimal time intervals 
we get $t_{i}-t_{i-1}\rightarrow dt$ and the sum becomes an integral, 
\begin{equation}
\label{eq:het-int}
C_{\vec{r}}(t,t')=g\left[\int_{t'}^{t}\frac{dt''}{\tau_{\vec{r}}(t'')}\right].
\end{equation}
In this general expression for $C_{\vec{r}}(t,t')$, 
the relaxation is allowed to be locally non-exponential, and the 
relaxation time is allowed to fluctuate both in space and time.

Another way of looking at local correlations is to consider first the exponential 
relaxation case where
\begin{equation}
\frac{dC(t,t')}{dt}=-\frac{1}{\tau}C(t,t')\qquad\Rightarrow C(t,t')=
   C_0\exp\left(-\frac{t-t'}{\tau}\right).
\end{equation}
If we allow $\tau$ to fluctuate in space and time but postulate 
that the same differential equation still holds, then we have 
\begin{equation}
\label{eq:het-2}
\frac{dC_{\vec{r}}(t,t')}{dt}=-\frac{1}{\tau_{\vec{r}}(t)}C_{\vec{r}}(t,t')\qquad\Rightarrow
C_{\vec{r}}(t,t')=C_0\exp\left[-\int^{t}_{t'}\frac{dt''}{\tau_{\vec{r}}(t'')}\right],
\end{equation}
where $C_{\vec{r}}(t,t')$ and $\tau_{\vec{r}}(t)$ are defined as
before and we have used the initial condition $C_0=\lim_{t \to
  t'^{+}}C(t,t')$.  Eq.~(\ref{eq:het-2}) can be extended to describe
the more general case of non-exponential relaxation. For example, we
can modify the right hand-side of the differential equation in
Eq.~(\ref{eq:het-2}) to
$\frac{dC_{\vec{r}}(t,t')}{dt}=-\frac{1}{\tau_{\vec{r}}}G[C_{\vec{r}}(t,t')]$.
Here the function $G[C_{\vec{r}}(t,t')]$ represents the effect of
non-exponential relaxation.  The solution of the new differential
equation, which represents the long-time limit of the local
correlation function, is given by
\begin{equation}
\label{eq:het-gen}
C_{\vec{r}}(t,t')={\cal C}\left[\phi_{\vec{r}}(t)-\phi_{\vec{r}}(t')\right], 
\end{equation}
where $\phi_{\vec{r}}(t)\equiv \int^{t}\frac{dt'}{\tau_{\vec{r}}(t')}$, 
and ${\cal C}(x)$ is a monotonous decreasing function that satisfies 
${\cal C}'(x)=-G[{\cal C}(x)]$ and $0\leq{\cal C}(x)\leq 1$.

By comparing Eq.~(\ref{eq:het-int}) with Eq.~(\ref{eq:het-gen}) we
find that $g(x)={\cal C}(x)$ and
\begin{equation}
\label{eq:phi-to-tau}
\phi_{\vec{r}}(t)-\phi_{\vec{r}}(t')=\int_{t'}^{t}
           \frac{dt''}{\tau_{\vec{r}}(t'')}.
\end{equation}
A completely different way of deriving Eq.~(\ref{eq:het-gen}) is based
on the presence of time reparametrization symmetry, and we describe it
next.  Time reparametrization symmetry refers to an invariance under
transformations of the time variable $t\rightarrow\phi(t)$, where
$\phi(t)$ is a continuous and monotonically increasing function of
$t$.  This symmetry has been shown to be present in the long-time
dynamics of mean field spin glass models~\cite
{Cugliandolo-Kurchan-jphysa94,Cugliandolo-Kurchan-prl-93} and of short
range spin glass models~\cite{Chamon2002,
  Castillo2003,Castillo2008,Mavimbela2011}.  Numerical studies in spin
glasses~\cite{Castillo2003} and structural glasses~\cite{Castillo2007,
  Avila2011} have given strong evidence supporting the presence of
this symmetry. The symmetry has also been used, from a different point
of view, to determine properties of symmetric observables whose full
dynamical properties are hard to
compute~\cite{Parisi-Franz-lett,Parisi-Franz-epje}. The symmetry is
spontaneously broken by correlations and relaxation functions.  We can
illustrate the breaking of the symmetry by considering the correlation
function $C(t-t')$. If $C(t-t')$ was symmetric under time
reparametrizations $t\rightarrow\phi(t)$, then
$C(t-t')=C[\phi(t)-\phi(t')]$ for {\em all} mappings
$t\rightarrow\phi(t)$, which is only possible if the correlation is
independent of time. But we know that in the long time limit the
correlation function of a glass forming liquid decays from its plateau
value, therefore the symmetry is broken by $C(t-t')$.  The broken
reparametrization symmetry is expected to give rise to Goldstone
modes~\cite{Chamon2002, Castillo2002, Castillo2003, Castillo2007,
  Castillo2008, Parsaeian2008a, Parsaeian2008, Parsaeian2009,
  Mavimbela2011, Avila2011,triangular_long} associated to smooth
variations in space of the reparametrized variable, that is $t \to
\phi_{\vec{r}}(t)$. The spatio-temporal fluctuations of
$\phi_{\vec{r}}(t)$ imply that different regions of a sample relax
differently from each other, that the relaxation of the different
regions is advanced (or retarded) with respect to each other and the
bulk, and that "advanced" and "retarded" regions can switch roles in
the course of the relaxation.  Hence, the spatio-temporal fluctuations
induced by the broken reparametrization symmetry provide a possible
explanation for dynamical heterogeneities in glass forming and glassy
systems. In addition, the broken reparametrization symmetry leads to
recovering Eq.~(\ref{eq:het-gen}) if we ignore longitudinal
fluctuations~\cite{Chamon2002, Castillo2002, Castillo2003}.
Longitudinal fluctuations should be suppressed by coarse graining in
larger regions because at long distances they are less correlated than
transverse fluctuations~\cite{Zwerger2004}.  Finally, it should be
noticed that there is a gauge symmetry under time-independent shifts
of the reparametrization variable $\phi_{\vec{r}}(t)$, that is in
Eq.~(\ref{eq:het-gen}), the local correlation is unchanged by the
transformation $\phi_{\vec{r}}(t) \to \phi_{\vec{r}}(t) +
\rho_{\vec{r}}$.  Consequently, when we analyze data we cannot
determine an absolute $\phi_{\vec{r}}(t)$.  To work with
gauge-invariant quantities, we either have to go back to studying
two-time quantities like $\phi_{\vec{r}}(t)-\phi_{\vec{r}}(t')$ or we
must work with time derivatives of $\phi_{\vec{r}}(t)$.

By analogy to the spin glass case, we expect that time
reparametrization symmetry, if present, can only be exact in the long
time limit, and that at any finite timescale there are symmetry
breaking terms, which are expected to become weaker at longer
times. In particular, if there is a finite relaxation time in the
system, it will provide a cutoff for long timescales, thus precluding
the symmetry to show its full effect. By contrast, for lower
temperatures, the dynamics at longer timescales can be explored, and
the effects of the symmetry could in principle manifest themselves
more clearly. Based on this discussion, we want to test the hypothesis
that at low temperatures and for large coarse-graining regions, the
long time correlations are well described by
Eq.~(\ref{eq:het-gen}). This is equivalent to the statement that
transverse fluctuations (Goldstone modes) are much stronger than
longitudinal fluctuations. This interplay between temperature and
region sizes implies that the crossover from dominant longitudinal
fluctuation to dominant transverse fluctuations should occur for
smaller coarse-graining regions for lower temperatures and for larger
coarse-graining regions for higher temperatures. In particular, at
temperatures moderately below the mode coupling critical temperature
$T_c$, we expect that for long timescales the coarse graining size for
which the transverse fluctuations start to be dominant will be smaller
than at temperatures moderately above $T_c$.

\section{Method to extract phases and instantaneous relaxation rates
  from numerical or experimental data}
\label{sec:method}

Our method for the extraction of $\phi_{\vec{r}}(t)$ from local two-time correlations is 
inspired by the work in Refs.~\cite{Avila2011,triangular_long}. 
The authors of Refs.~\cite{Avila2011,triangular_long}
use a functional form $g_{\rm global}(x)=q_{EA}\exp\left(-|x|^{\beta}\right)$, where $q_{EA}$ and  
$\beta$ are fitting parameters, to fit global correlations 
in molecular dynamics simulations of particle systems and polymer systems. They 
find functions $\phi(t)$ for the different systems such that the data are described by 
\begin{equation}
C(t,t')=g_{\rm global}[\phi(t)-\phi(t')],
\label{eq:global-fit}
\end{equation}
As discussed in Sec.~\ref{sec:equivalence}, local fluctuations can be
described in terms of Goldstone modes such that  
\begin{equation}
C_{\vec{r}}(t,t')\approx g[\phi_{\vec{r}}(t)-\phi_{\vec{r}}(t')].
\label{eq:local-fit}
\end{equation}
In principle, the functions $g_{\rm global}(x)$ and $g(x)$,
respectively describing global two-time correlations and local
two-time correlations, could have completely different forms. For
example, as discussed in Sec.~\ref{sec:equivalence}, some of the
initial motivation for the picture of dynamical heterogeneity came
from the idea that non-exponential relaxation in a macroscopic sample
is due to the combined effect of local exponential relaxations with
different relaxation times. In this picture, $g_{\rm global}(x)
\propto \exp(-|x|^{\beta})$ and $g(x) \propto \exp(-|x|)$. In
Refs.~\cite{Avila2011,triangular_long}, as a simplifying assumption, the two functions
were taken to be equal. In the present work, we focus exclusively on
the local function $g(x)$. However, we will consider different coarse
graining sizes when determining $C_{\vec{r}}(t,t')$. Each of these
coarse graining sizes will give rise to a different form for
$g(x)$. In practice, to simplify the determination of $g(x)$, we will
restrict it to be a member of a family of functional forms
$g(x;\vec{\alpha})$, parametrized by a vector $\vec{\alpha}$ of $p$
components, and for each coarse graining size both the parameter
vector $\vec{\alpha}$ and the fluctuating phases $\phi_{\vec{r}}(t)$ will be
determined by fitting $C_{\vec{r}}(t,t')$ according to
Eq.~(\ref{eq:local-fit}). For technical reasons, we impose the
condition that $g$ be an even function, $g(-x) = g(x)$. 

Let's consider a data set containing "snapshots" of the relevant
degrees of freedom of the system, taken at times
$\{t_i\}_{i=1,...,M}$. In our case the positions of the particles are
recorded at each time step.
For a given coarse graining region centered at a point $\vec{r}$ in
the sample, our data points will be the $M(M-1)/2$ two-time local
correlations $C_{\vec{r}}(t_i,t_j)$ calculated from the recorded
data. Considering for the moment a fixed parameter vector
$\vec{\alpha}$, we have $M$ fitting parameters
$\{\phi_{\vec{r}}(t_i)\}_{i=1,...,M}$, thus giving us a fitting
problem with $\sim M/2$ data points per fitting parameter. However,
since the function $g(x)$ is nonlinear, this becomes a nonlinear
fitting problem with a large number $M$ of fitting parameters. In
order to make this problem manageable, we want to convert it into a
linear fitting problem. To do this, we expand the rhs of
Eq.~(\ref{eq:local-fit}) to linear order in the local fluctuations
\begin{equation}
  g[\phi_{\vec{r}}(t)-\phi_{\vec{r}}(t')] \approx g[\phi(t) - \phi(t')] +
  g'[\phi(t) - \phi(t')]
  [\delta\phi_{\vec{r}}(t)-\delta\phi_{\vec{r}}(t')]
\label{eq:g_r_linear}
\end{equation}
where $\phi(t)$ is a global phase, still to be determined, and the
local fluctuations of the phase are given by
\begin{equation}
  \delta\phi_{\vec{r}}(t)\equiv\phi_{\vec{r}}(t)-\phi(t).
\label{eq:dphi_def}
\end{equation}
At this point, we have simplified the problem by converting
it into a linear fitting problem, at the price of introducing the
extra phase variables $\{\phi(t_i)\}_{i=1,\cdots,M}$. 

Before we describe the method for determining the phases, we define
three quantities to measure fluctuations in the two-time local
correlations. 
The {\em total fluctuations} are defined by
\begin{equation}
\delta C_{\vec{r}}(t,t') \equiv
C_{\vec{r}}(t,t')-g[\phi(t)-\phi(t')]
\label{eq:dC-def}
\end{equation}
and contain complete information about the space dependence of the
local two time correlations. We decompose the total fluctuations as the
sum of a transverse component $\delta C^{T}_{\vec{r}}(t,t')$ and a
longitudinal component  $\delta C^{L}_{\vec{r}}(t,t')$,  
\begin{equation}
\delta C_{\vec{r}}(t,t')=\delta C^{T}_{\vec{r}}(t,t')+\delta
C^{L}_{\vec{r}}(t,t').
\label{eq:dC-T+L}
\end{equation}
The transverse component is associated to time reparametrization
fluctuations, and it is defined by the linear term in
Eq.~(\ref{eq:g_r_linear}), namely
\begin{equation}
\delta C^{T}_{\vec{r}}(t,t') \equiv g'[\phi(t)-\phi(t')]
\left[\delta\phi_{\vec{r}}(t)-\delta\phi_{\vec{r}}(t')\right]. 
\label{eq:dCT-def}
\end{equation}
To satisfy Eq.~(\ref{eq:dC-T+L}), the longitudinal component is defined by 
\begin{eqnarray}
\delta C^{L}_{\vec{r}}(t,t') & \equiv & \delta
C_{\vec{r}}(t,t')-\delta C^{T}_{\vec{r}}(t,t') \\
& \equiv & C_{\vec{r}}(t,t')- \left\{ g[\phi(t)-\phi(t')] + g'[\phi(t)-\phi(t')]
\left[\delta\phi_{\vec{r}}(t)-\delta\phi_{\vec{r}}(t')\right] 
\right\}. 
\label{eq:dCL-def}
\end{eqnarray}
In what follows we explain how to perform a least squares fit of the
local correlation $C_{\vec{r}}(t,t')$ by the rhs of
Eq.~(\ref{eq:g_r_linear}). In the case of numerical data, we normally
have not only data for different coarse grained regions inside the
sample, but also data for different independent runs. If we have $N_R$
independent simulation runs for a system of volume $V$, we can imagine
juxtaposing the configurations for all the runs, thus creating a
larger volume ${\cal V} = N_R V$. Assuming that each coarse graining
region $B_{\vec{r}}$ centered around point $\vec{r}$ has a volume
$V_{cg}$, the total number of non-overlapping regions per run is
$\omega = [V/V_{cg}]$, and the total number of non-overlapping regions
including all runs is $\Omega = N_R \omega = N_R [V/V_{cg}]$, where in this
context $[V/V_{cg}]$ denotes the integer part of $V/V_{cg}$.

The fit we want to perform corresponds to minimizing the quantity
\begin{equation}
E \equiv \frac{1}{\Omega} \sum_{\vec{r}}
\epsilon\left[\{\delta\phi_{\vec{r}}(t_i)\}_{i=1,\cdots,M};\vec{\alpha},
\{C_{\vec{r}}(t_j,t_i)\}_{1\le i<j \le M}\right]
\end{equation}
with respect to all phase fluctuations and with respect to
$\vec{\alpha}$. Here
\begin{eqnarray}
& & \epsilon\left[\{\delta\phi_{\vec{r}}(t_i)\}_{i=1,\cdots,M};
  \vec{\alpha}, \{C_{\vec{r}}(t_j,t_i)\}_{1\le i<j \le M}\right] \equiv
  \eta^{-1}(M) \sum_{1\le i<j \le M} \left[\delta
    C^{L}_{\vec{r}}(t_j,t_i) \right]^2,\\ 
& & \qquad \qquad = \eta^{-1}(M) \sum_{1\le i<j \le M} \left(
    C_{\vec{r}}(t,t_i)- \left\{ g[\phi(t_j)-\phi(t_i);\vec{\alpha}] +
    g'[\phi(t_j)-\phi(t_i);\vec{\alpha}]
    \left[\delta\phi_{\vec{r}}(t_j)-\delta\phi_{\vec{r}}(t_i)\right]
    \right\} \right)^2, 
  \label{eq:epsilon-C-Taylor}
\end{eqnarray}
is the residual corresponding to the region centered at $\vec{r}$, 
\begin{equation}
\eta(M) \equiv \frac{M(M-1)}{2} - M + 1 = \frac{M(M-3)+2}{2}
\end{equation} 
is the number of degrees of freedom in the fit of the data
corresponding to one particular coarse graining region, and
$\sum_{\vec{r}}$ denotes a sum over non-overlapping coarse graining
regions. From now on we consider only times belonging to the set
$\{t_i\}_{i=1,\cdots,M}$, and we simplify our notation by indicating
times as subindices, {\em i.e.\/} $\delta \phi_{i \vec{r}} \equiv \delta
\phi_{\vec{r}}(t_i)$, $\delta C^{L}_{ji \vec{r}}\equiv \delta
C^{L}_{\vec{r}}(t_j,t_i)$, and similarly for all other one- and
two-time variables.

To minimize $E$, we observe that for a fixed parameter vector
$\vec{\alpha}$, the determination of the local fluctuations
$\{\delta\phi_{i \vec{r}}\}$ for a specific region centered at
$\vec{r}$ can be performed by separately minimizing the corresponding
$\epsilon\left(\{\delta\phi_{i \vec{r}}\}_{i=1,\cdots,M};\vec{\alpha},
\{C_{ji \vec{r}}\}_{1\le i<j \le M}\right)$ for that particular
region. For this reason, instead of minimizing with respect to all
variables in one step, the problem becomes significantly simplified if
we perform an iterated minimization,
\begin{eqnarray}
\min_{ \{\delta\phi\}; \vec{\alpha} } E & = &
\min_{ \vec{\alpha} } \epsilon_{\phi}(\vec{\alpha}), \\
\epsilon_{\phi}(\vec{\alpha}) & \equiv &
\frac{1}{\Omega} \sum_{\vec{r}} \min_{\{\delta\phi_{i \vec{r}}\}}
\epsilon\left(\{\delta\phi_{i \vec{r}}\};\vec{\alpha},\{C_{ji \vec{r}}\}_{1\le i<j \le M}\right).
\end{eqnarray}
In other words, we first keep $\vec{\alpha}$ fixed and separately
minimize with respect to the phase fluctuations in each coarse
graining region, and then minimize the result with respect to
$\vec{\alpha}$.  The minimization with respect to the phases, at fixed
$\vec{\alpha}$, is performed in two steps. In the first step we
determine global phases. In the second step, we use the global phases
from the first step to determine local phases. Next, we describe these
steps in detail, but without derivations. The details of the
derivations can be found in Appendix~\ref{app:matrix-method}. 

\subsection{Step One: determining global phases at fixed $\alpha$}
The global phases $\{\phi_i\}_{i=1,...,M}$ define the point around
which the Taylor expansion in Eq.~(\ref{eq:g_r_linear}) is
performed. The only requirement on them is to be close enough to the
values of $\phi_{i \vec{r}}$ such that the expansion is accurate to
first order in the fluctuations. In particular, we only need to
determine them to within an error of the order of the fluctuations in
the $\phi_{i \vec{r}}$. There is more than one possible way to satisfy
these conditions, and for simplicity we do it by choosing them to be
the phases that best represent the {\em global} correlations for the
given value of $\vec{\alpha}$.  We define the quantity
$\bar{\epsilon}$
\begin{equation}
\bar{\epsilon}(\{\phi_1,\cdots,\phi_M\};\vec{\alpha}) \equiv
\eta^{-1}(M) \sum_{1 \le i < j \le M} \left[C_{ji} -
g^{(1)}(\phi_j-\phi_i;\vec{\alpha})\right]^2, 
\label{eq:global-epsilon}  
\end{equation}
where $C_{ji} \equiv C(t_j,t_i)$, and
$g^{(1)}(\phi_j-\phi_i;\vec{\alpha})$ is the first order Taylor
expansion of $g(\phi_j-\phi_i;\vec{\alpha})$ with respect to the phase
difference, taken about
\begin{equation}
\Phi_{ji}\equiv \left[\sgn(t_j-t_i)\right]
g^{-1}\left[C(t_j,t_i)\right].  
\label{eq:Phi-def}
\end{equation}
To minimize $\bar{\epsilon}$ we impose the condition that all
derivatives of $\bar{\epsilon}$ with respect to $\phi_k$ must be
zero. Additionally, we fix the gauge with the condition
\begin{equation}
0 = \sum_{i=1}^M \phi_i.
\label{eq:g-gauge-cond}
\end{equation}
Thus we obtain the matrix equation 
\begin{equation}
\vec{\bar{w}} = \bar{A} \; \vec{\phi}
\label{eq:global-phase}
\end{equation} 
where 
\begin{eqnarray}
\bar{w}_k & \equiv & \left\{ 
\begin{array}{l@{\quad\mbox{for}\quad}l}
0 & k=1 \\
\sum_{j=1}^M  \Phi_{jk} \; g'^2(\Phi_{jk};\vec{\alpha}) & k \ne 1
\end{array} 
\right.
\\
\bar{A}_{ki} & \equiv & \left\{ 
\begin{array}{l@{\quad\mbox{for}\quad}l}
1 & k=1 \\
-\delta_{ki}\sum_{j=1}^M g'^2(\Phi_{jk};\vec{\alpha})
+ g'^2(\Phi_{ki};\vec{\alpha})
& k \ne 1.
\end{array} 
\right.
\label{eq:bar-w-bar-A}
\end{eqnarray}
The  solution to Eq.~(\ref{eq:global-phase}) is the set of global phases 
$\{\phi_i\}_{i=1,...,M}$ which minimizes $\bar{\epsilon}$ for the given 
$\vec{\alpha}$. We use the global phases as input in the second step, 
which we describe next.

\subsection{Step Two: determining the local phases at fixed $\alpha$}
In the second step, we minimize $\epsilon(\{\delta\phi_{i
  \vec{r}}\};\vec{\alpha})$ with respect to the local phase
fluctuations $\{ \delta\phi_{1 \vec{r}}, \cdots, \delta\phi_{M
  \vec{r}} \}$, for each coarse graining region, while keeping
$\vec{\alpha}$ fixed at the values used in the first step.  By Taylor
expanding $g(\phi_{i\vec{r}}-\phi_{j\vec{r}};\vec{\alpha})$ about
the global phase differences, $\phi_i-\phi_j$, imposing the condition
that all derivatives of $\epsilon(\{\delta\phi_{i
  \vec{r}}\};\vec{\alpha})$ with respect to the phase fluctuations $\{
\delta\phi_{1 \vec{r}}, \cdots, \delta\phi_{M \vec{r}} \}$ should be
zero, and fixing the gauge with the condition
\begin{equation}
0 = \sum_{i=1}^M \delta \phi_{i \vec{r}}, 
\label{eq:l-gauge-cond}
\end{equation}
we obtain the matrix equation 
\begin{equation}
\vec{w}_{\vec{r}}= A \; \delta\vec{\phi}_{\vec{r}},
\label{eq:local-phase}
\end{equation}
where 
\begin{eqnarray}
w_{k \vec{r}} & \equiv & \left\{ 
\begin{array}{l@{\quad\mbox{for}\quad}l}
0 & k=1 \\
\sum_{j=1}^M g'(\phi_k-\phi_j;\vec{\alpha}) \; \delta
C_{kj\vec{r}} & k \ne 1
\end{array} 
\right.
\\
A_{ki} & \equiv & \left\{ 
\begin{array}{l@{\quad\mbox{for}\quad}l}
1 & k=1 \\
\delta_{ki}\sum_{j=1}^M g'^2(\phi_k-\phi_j;\vec{\alpha})
- g'^2(\phi_k-\phi_i;\vec{\alpha})
& k \ne 1.
\end{array} 
\right.
\label{eq:w-A-local-def}
\end{eqnarray}
The solution to this matrix equation is the vector
$\delta\vec{\phi}_{\vec{r}} = (\delta\phi_{1 \vec{r}}, \cdots,
\delta\phi_{M \vec{r}})$ containing the local fluctuating phase
differences that minimize $\epsilon$ for the given value of the
parameter vector $\vec{\alpha}$ and for the region $B_{\vec{r}}$. By
averaging the minimum values of $\epsilon(\{\delta\phi_{i
  \vec{r}}\};\vec{\alpha})$ for all regions, we obtain
$\epsilon_{\phi}(\vec{\alpha})$.

\subsection{Minimization with respect to $\alpha$}
The optimum value of $\vec{\alpha}$ is obtained by numerically
minimizing $\epsilon_{\phi}(\vec{\alpha})$ with respect to it. Once
the optimum $\vec{\alpha}$ has been obtained, the solutions of
Eq.~(\ref{eq:local-phase}) for that specific value of $\vec{\alpha}$
and for each region $B_r$ give the optimal choice for the fluctuating
phases $\delta\phi_{i \vec{r}}$. 

\section{Testing the method with data from numerical simulations of
  glass-forming model systems}
\label{sec:tests}

We tested the method on data from classical Molecular Dynamics
simulations of two glass forming systems, each one consisting of an
$80:20$ binary mixture of $1000$ particles interacting via purely
repulsive Weeks-Chandler-Andersen (WCA)
potentials~\cite{Parsaeian2009,fnote}
An initial configuration of the systems was prepared by randomly displacing 
the particles from a simple cubic lattice configuration and assigning each 
particle a velocity taken from a Maxwell-Boltzmann distribution, with the 
distribution corresponding to the simulation's temperature. Next, the
systems were equilibrated, by integrating the velocity Verlet 
algorithm, at an initial temperature $T_i=5.0$ much 
higher than the Mode Coupling critical temperature $T_c=0.263\pm
0.01$ and then instantaneously quenched to a final temperature~\cite{Parsaeian2009}. 
In~\cite{Parsaeian2009}, $T_c$ was
determined by measuring the $\alpha$-relaxation time $\tau_{\alpha}(T)$
and fitting it with a power law as a function of temperature,
$\tau_{\alpha}(T)=\tau_0\left(\frac{T_c}{T-T_c}\right)^\gamma$, with
$\gamma=2.1\pm 0.7$. One of the systems was quenched to a temperature
$T = 0.23 \sim 0.9 T_c$ and it did not reach equilibrium during the
simulation. We call this system the {\it aging system}. The other
system was quenched to a temperature $T=0.29 \sim 1.1 T_c$, and it
reached equilibrium within the duration of the simulation. In our
studies we use the data corresponding to times after the system
reached equilibrium and we call this system the {\it equilibrium
  system}.  After the quench, the systems were allowed to evolve for
times of order $10^{-8}\quad s$, which are much longer than the typical vibrational periods of the
particles. The simulations were done at constant particle number, volume 
and temperature. 
To keep the temperature constant, the velocities of the 
particles were periodically rescaled so that the kinetic energy of the 
system equals $\frac{3}{2}Nk_BT$. 
Snapshots of the systems were taken periodically during
the evolution and the positions of the particles were stored.

\begin{figure}[h]
  \centerline{
     \includegraphics[width=6cm,angle=270]{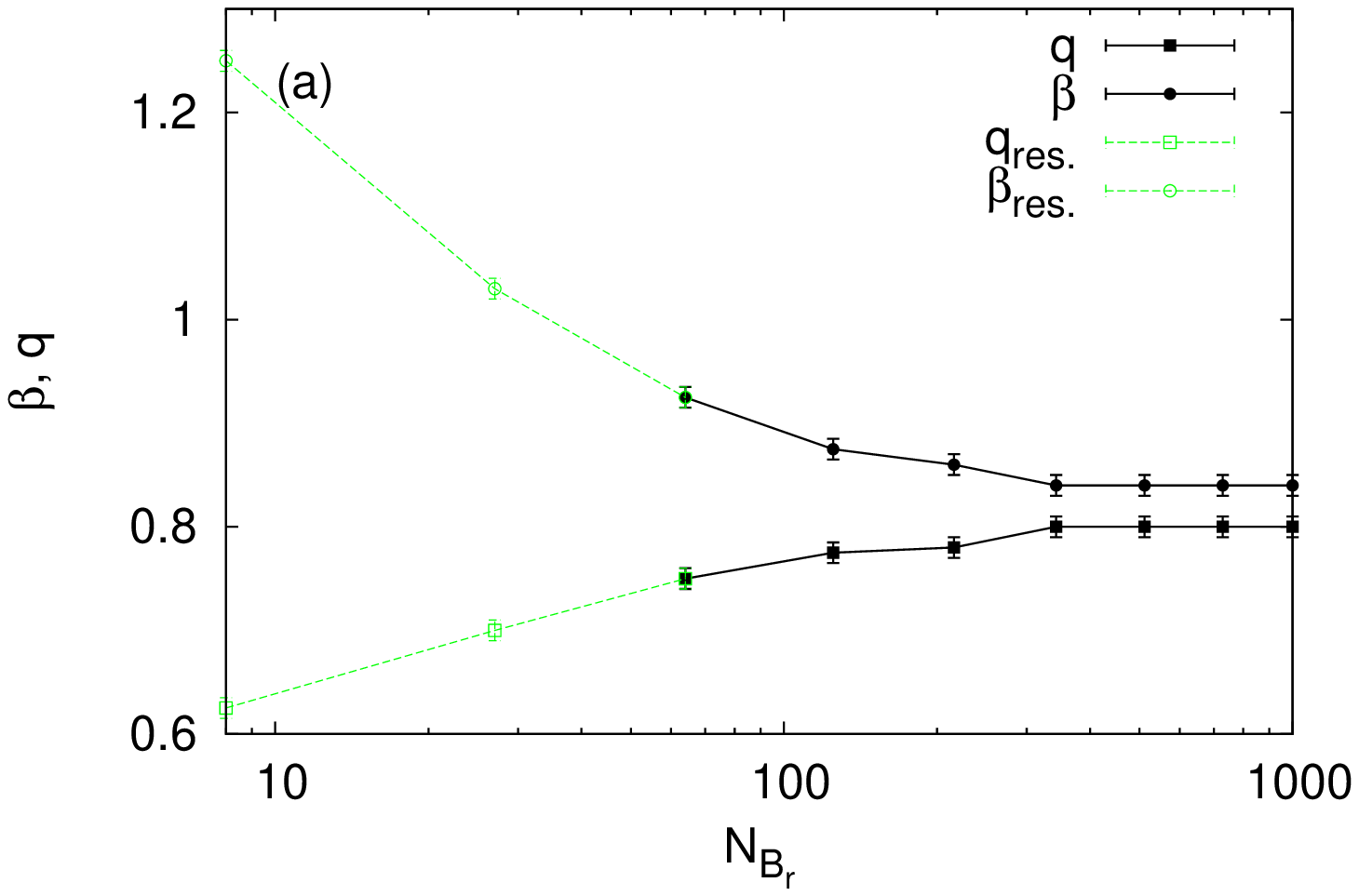}
     \hspace{0.3cm}
     \includegraphics[width=6cm,angle=270]{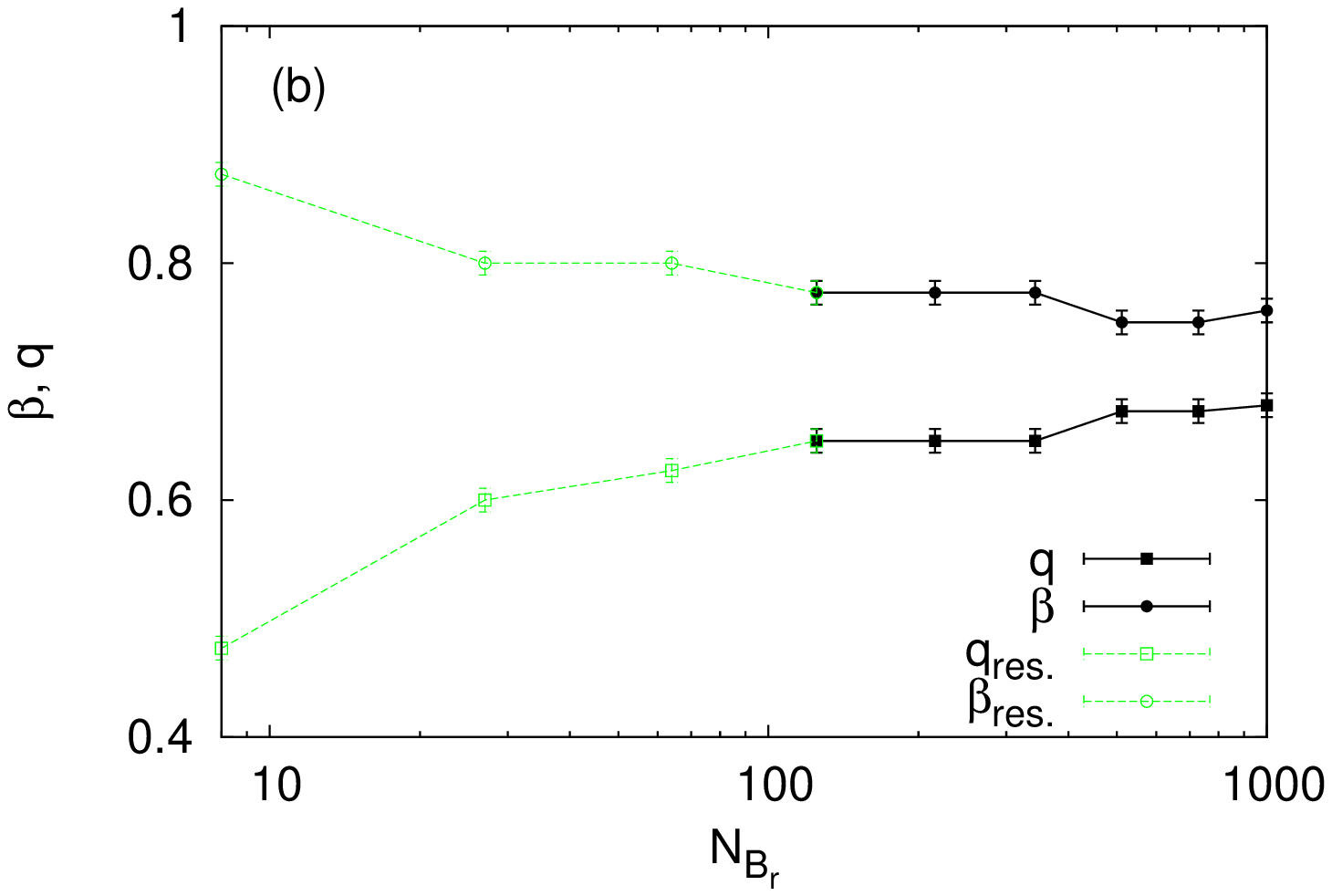}
      }
  \caption{(Color online) Variation of the optimal values for the
    fitting parameters $\beta$ and $q_{EA}$ with coarse graining size,
    in the equilibrium system with $M=94$ snapshots and $N_R=10$
    independent runs (right panel), and in the aging system with
    $M=100$ and $N_R=10$ (left panel).  The part in black (solid
    squares $q$ and solid circles $\beta$) for each curve corresponds
    to the range of coarse graining sizes for which longitudinal
    fluctuations account for less than 40\% of the total
    fluctuations. The part in green (hollow squares $q_{res.}$ and
    hollow circles $\beta_{res.}$) for each curve corresponds to the
    range of coarse graining sizes for which longitudinal fluctuations
    account for more than 40\% of the total fluctuations. }
  \label{fig:qbeta-vc}
\end{figure}

To probe a system, we divide it into coarse graining boxes
$B_{\vec{r}}$, centered at the space points $\vec{r}$. The local
correlation for each box is defined by~\cite{Castillo2007}
\begin{equation}
 C_{\vec{r}}(t,t') \equiv \frac{1}{N_{B_{\vec{r}}}}\sum_{\vec{r}_i(t')\in B_{\vec{r}}}
 \cos\{\vec{q}\cdot[\vec{r}_i(t)-\vec{r}_i(t')]\}.
\end{equation}
Here $N_{B_{\vec{r}}}$ is the number of particles in the coarse
graining box $B_{\vec{r}}$ at time $t'$, and $|\vec{q}|$ is chosen to
be the value given by the location of the main peak of the static
structure factor $S(q)$, {\em i.e.\/} $q=7.2$. $C_{\vec{r}}(t,t')$
measures the extent to which the configuration in the coarse graining
region has changed between the times $t'$ and $t$. If the particles
initially in the region have not moved much then $C_{\vec{r}}(t,t')$
is larger than $1/2$, and if the particles have moved a significant
distance then $C_{\vec{r}}(t,t') \ll 1$. In the first case, we say the
the region is ``slow'', and in the second, that it is ``fast''. To
implement the method, we choose the fit function
$g\left(x;\vec{\alpha}\right)$ to be a stretched exponential
\begin{equation} 
g\left(x;\vec{\alpha}\right) = q_{EA} \exp\left(-|x|^{\beta}\right).
\end{equation}
In this case the vector of parameters is $\vec{\alpha} \equiv
(q_{EA},\beta)$, where $q_{EA}$ is the plateau value of the
correlation function and $\beta$ is the stretching exponent. We begin
the analysis by determining the optimal values of $\vec{\alpha}$ at
different coarse graining sizes; the results are summarized in
Fig.~(\ref{fig:qbeta-vc}) for both systems.
Each one of the curves is drawn using two colors: one part is drawn in black
and the other in a color other than black. The part that is
drawn in black is the part where longitudinal fluctuations, {\em
  i.e.\/} the fit residuals, account for more than 40\% of the total
variance of the fluctuations. We observe that the optimal $\beta$
decreases with increasing coarse graining volume and the optimal
$q_{EA}$ increases with coarse graining volume. Both quantities
approach constant values at high coarse graining volumes. The trend on
$q_{EA}$ appears to be an artifact of the fact that fluctuations are
stronger for smaller coarse graining volumes. Since fluctuations that
increase $C_{\vec{r}}(t,t')$ above a value given by the Debye-Waller
factor are unlikely, there is a bias for fluctuations to reduce,
rather than increase, the value of $C_{\vec{r}}(t,t')$ for times when
$C_{\vec{r}}(t,t')$ is high. Since the largest values that
$C_{\vec{r}}(t,t')$ takes are associated to the optimal $q_{EA}$, this
bias will push $q_{EA}$ down, particularly when fluctuations are
stronger due to the small size of the coarse graining regions. The
trend on $\beta$ is reminiscent of one of the initial motivations for
the proposal that dynamics must be heterogeneous, {\em i.e.\/} the
idea~\cite{Sillescu1999, Ediger2000} that in systems for which the
bulk relaxation is non-exponential, local regions have exponential
relaxation - $\beta \approx 1$ - with heterogeneous timescales, and this
translates into a lower exponent $\beta$ for the system as a
whole. However, as we will discuss below, for smaller coarse graining
volumes the majority of the fluctuations are captured by the fit
residuals and not by the fitting function itself, so it is unclear
whether or not this trend in the optimal value of the fit parameter
$\beta$ has any physical significance.

\subsection{Fluctuations}
\label{sec:fluctuations}
As we mentioned in Sec.~\ref{sec:equivalence} we expect longitudinal
fluctuations to be less correlated at long distances, and consequently
to be more strongly suppressed by coarse graining, than transverse
fluctuations~\cite{Zwerger2004}. It is not immediately
obvious how to quantify the strength of the fluctuations, since they
are functions of the position and of two times. However, in this
context it becomes natural to think of them as vectors in an Euclidean
vector space, with the inner product defined by 
\begin{equation}
\left( f | g \right) \equiv \frac{2}{\omega M(M-1)}
\sum_{1 \le j < i \le M} \sum_{\vec{r}}  \langle 
f_{\vec{r}}(t_i,t_j) g_{\vec{r}}(t_i,t_j)\rangle,  
\label{eq:inner-product-def}
\end{equation}
and the Euclidean norm defined by
\begin{equation}
||f||\equiv \left( f|f \right)^{1/2}. 
\label{eq:mag-def}
\end{equation} 
Here $M$ is the number of snapshots, $\{ \vec{r} \}$ are the centers
of each one of the $\omega$ non-overlapping coarse-graining boxes
$B_{\vec{r}}$ in the volume $V$ of the system, and
$\langle\cdots\rangle$ denotes an average over thermal fluctuations,
which in the case of quantities obtained from numerical simulations
corresponds to an average over independent simulation runs. The only
slightly unusual aspect in these definitions is the prefactor
$2/\omega M(M-1)$, which was included so that the square of the norm
is an ``intensive'' quantity, which gives the average over positions
and over time pairs of the variance of the fluctuations for one coarse
graining box and one time pair. This makes it easier to compare
results for different snapshot numbers $M$ and different numbers
$\omega$ of coarse graining boxes. With these definitions, and as a
direct consequence of the minimization conditions that determine the
values of the phases $\{ \phi_{\vec{r}}(t_i) \}$, we show in
Appendix~\ref{app:orthogonality} that the transverse and longitudinal
fluctuation vectors are orthogonal, {\em i.e.\/}
\begin{equation}
\left( \delta C^{T} | \delta C^{L} \right) = 0, 
\label{eq:T-L-orthogonal}
\end{equation} 
and therefore they satisfy the Pythagorean condition
\begin{equation}
||\delta C||^2 = ||\delta C^{T}||^2 + ||\delta C^{L}||^2.
\label{eq:Pythagoras}
\end{equation}

\begin{figure}[h]
 \begin{minipage}[b]{0.5\linewidth}
  \centerline{ 
    \includegraphics[width=6cm,angle=270]{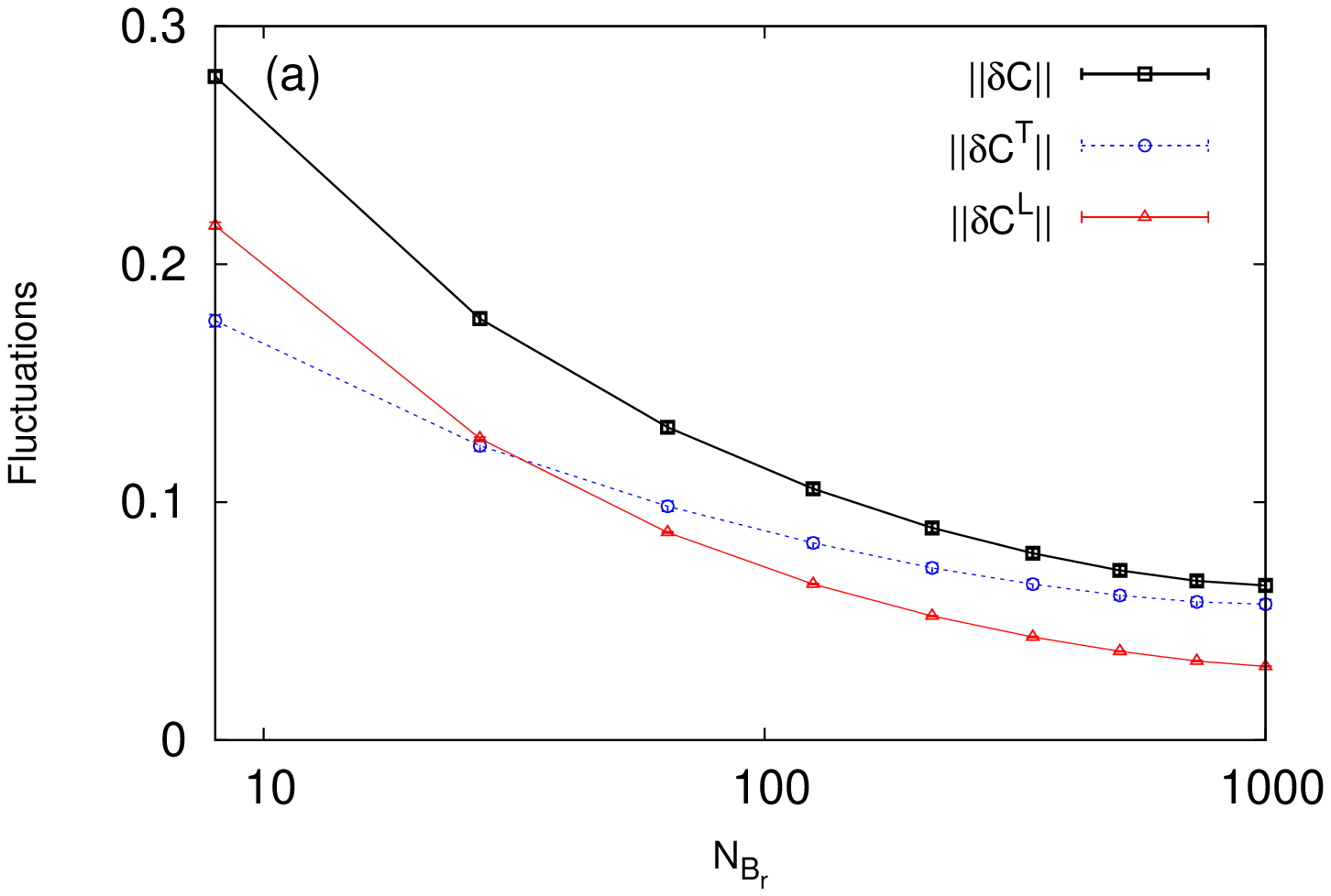} 
    \vspace{0.3cm} 
    \includegraphics[width=6cm,angle=270]{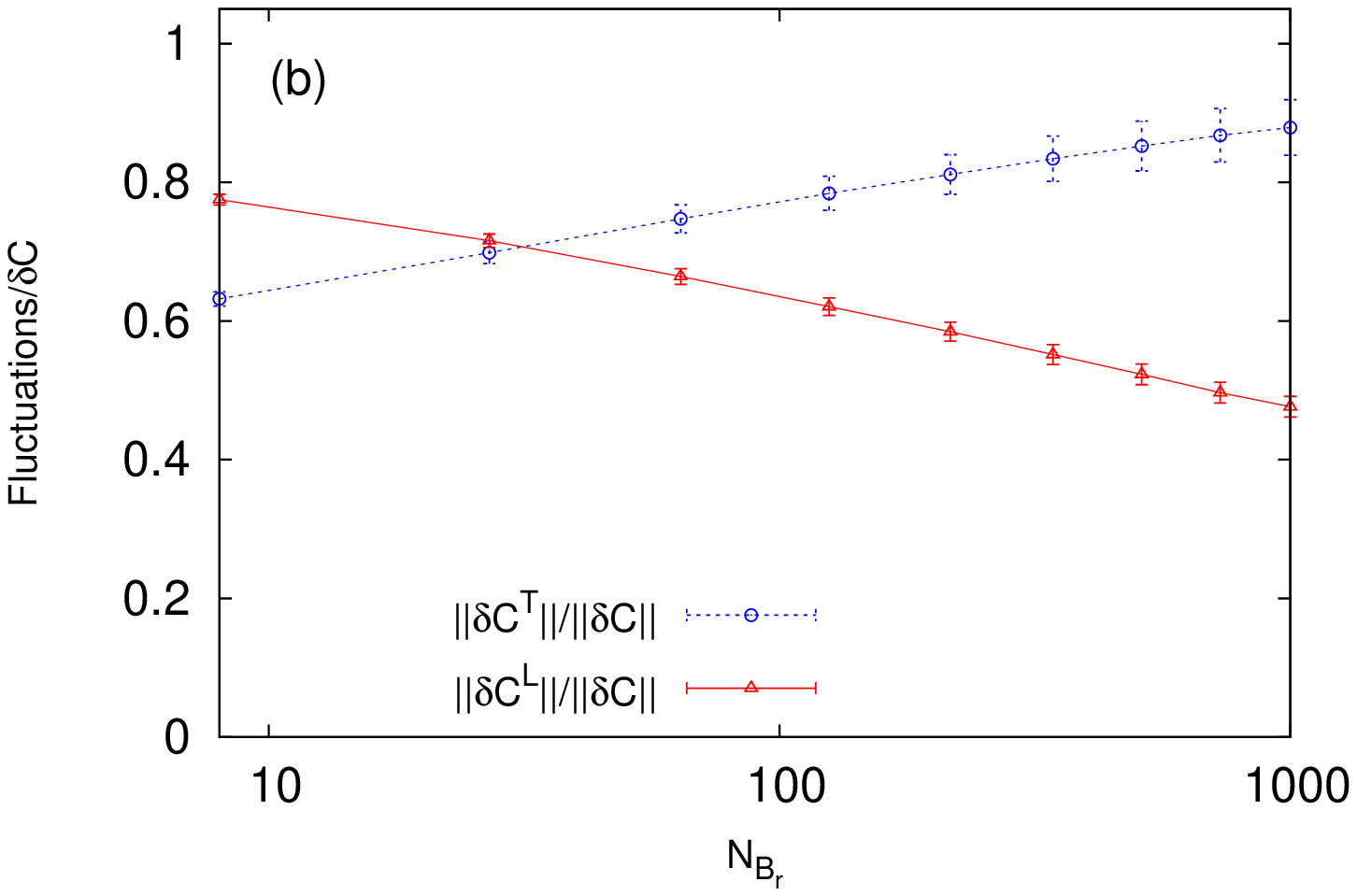}
  }
 \end{minipage}
    \hspace{0.5cm}
 \begin{minipage}[b]{0.5\linewidth}
   \centerline{
    \includegraphics[width=6cm,angle=270]{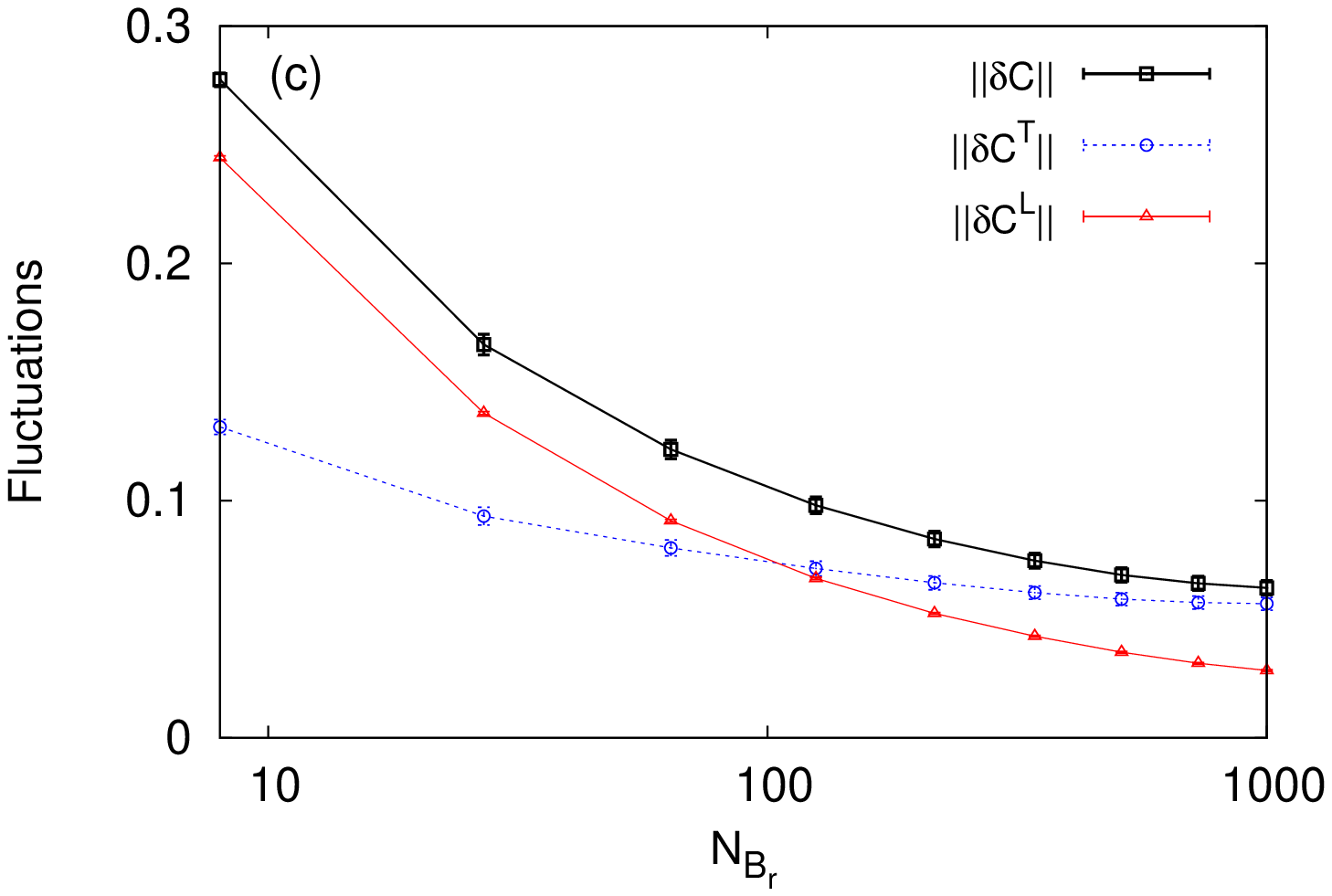}
    \vspace{0.3cm} 
    \includegraphics[width=6cm,angle=270]{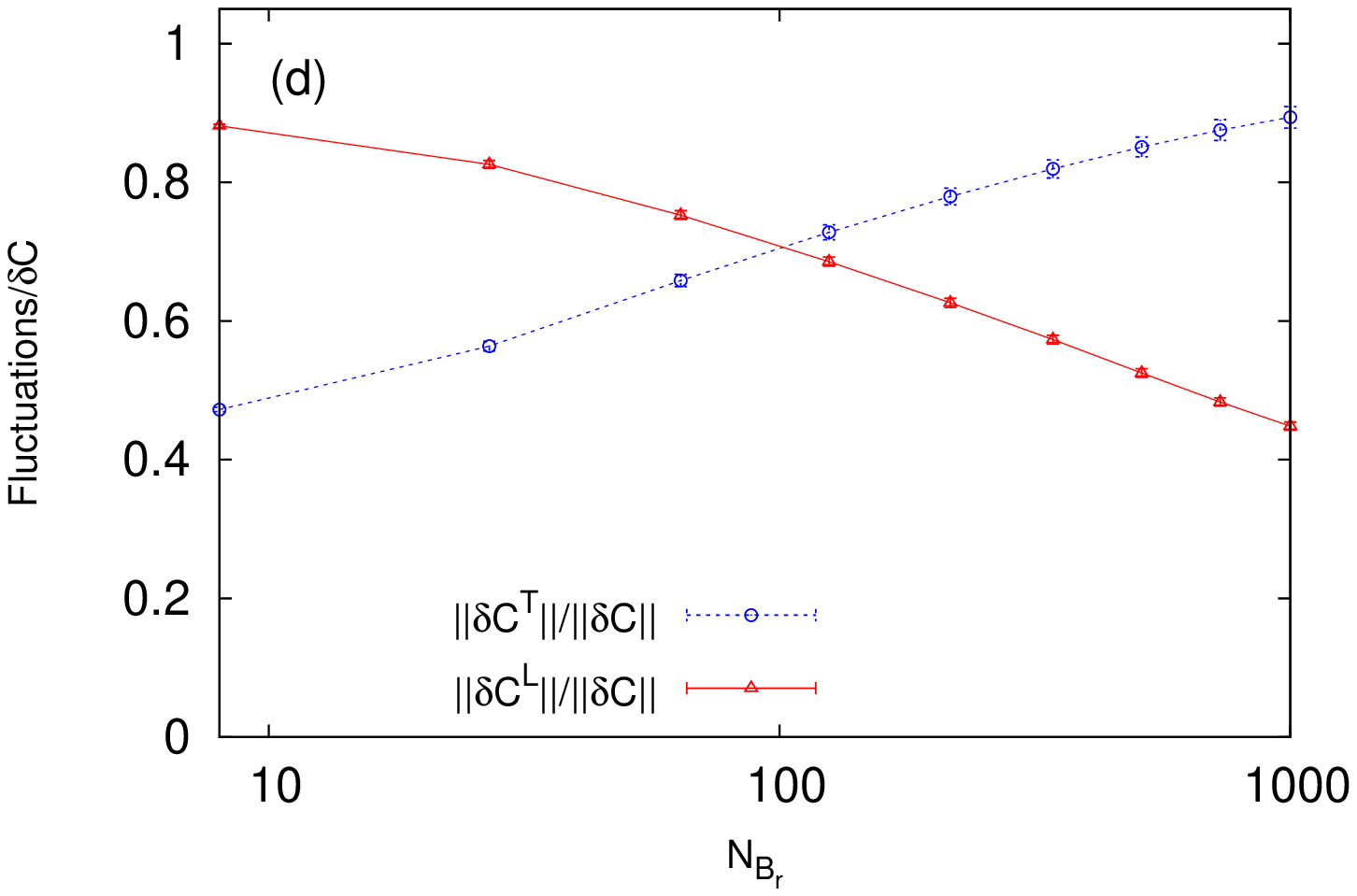}
   }
 \end{minipage}
  \caption{(Color online) Dependence of fluctuation strengths on the average number
    of particles $N_{B_{\vec{r}}}$ per coarse graining box, in the
    aging system with $M=100$ and $N_R=100$ (top panels) and in 
    the system in equilibrium with $M=94$ and $N_R=100$ (bottom
    panels). For each system, the left panel shows the magnitude
    $||\delta C||$ of the total fluctuations, the magnitude $||\delta
    C^{T}||$ of the transverse fluctuations, and the magnitude
    $||\delta C^{L}||$ of the longitudinal fluctuations; and the right
    panels show the ratios $||\delta C^{T}||/||\delta C||$ and
    $||\delta C^{L}||/||\delta C||$. The ratio $||\delta
    C^{T}||/||\delta C||$ increases and the ratio $||\delta
    C^{L}||/||\delta C||$ decreases with increasing coarse graining
    size. For both systems, the first ratio is smaller than the second
    one at small coarse grainings, and larger than the second one at
    large coarse grainings. The two ratios cross at $N_{B_{\vec{r}}}
    \approx 30$ for the aging system and at $N_{B_{\vec{r}}} \approx
    100$ for the equilibrium system. Error bars are shown, with some of 
    the bars not clearly visible because they are the same size as the 
    symbols or smaller.}
  \label{fig:fluct_v}
\end{figure}

In Fig.~(\ref{fig:fluct_v}) we show how the strengths of the
fluctuations vary with the average number of particles
$N_{B_{\vec{r}}}$ per coarse graining box. In the figure, the two top
panels correspond to the aging system and the two bottom ones
correspond to the equilibrium system.  For each system, the left
panel shows the magnitudes $||\delta C||$, $||\delta C^{T}||$, and
$||\delta C^{L}||$ of the total fluctuations, their transverse
component and their longitudinal component; and the right panel shows
the ratios $||\delta C^{T}||/||\delta C||$, and $||\delta
C^{L}||/||\delta C||$. Due to the Pythagorean condition, the sum of
the squares of these two ratios is unity in all cases. 

We expect that as the coarse graining is increased, all fluctuations
will be suppressed by the effect of averaging. Indeed, we find that
the magnitudes of all three kinds of fluctuations decrease as we
increase the coarse graining volume, both for the aging system and for
the equilibrium system. The magnitude of the longitudinal fluctuations
decreases at a faster rate than the total fluctuations while the
magnitude of the transverse fluctuations decreases at a slower
rate. Indeed at small coarse graining volumes the ratio $||\delta
C^{T}||/||\delta C||$ is smaller than the ratio $||\delta
C^{L}||/||\delta C||$, but the first one increases and the second one
decreases as the coarse graining volume increases. For the aging
system the two ratios cross at $N_{B_{\vec{r}}} \approx 30$ and for
the equilibrium system they cross at $N_{B_{\vec{r}}} \approx
100$. This is consistent with our expectation that longitudinal
fluctuations should be more strongly suppressed by coarse graining
than transverse fluctuations. We also observe that transverse
fluctuations are dominant for a wider range of coarse graining sizes
in the case of the aging system than in the case of the equilibrium
system. This is indeed what should be expected if the transverse
fluctuations are the Goldstone fluctuations associated with time
reparametrization symmetry. Since time reparametrization symmetry is a
long time asymptotic effect, and in the case of the equilibrium
system, which is at higher temperature than the aging system, the
relaxation time provides a cutoff for long timescales, we expect that
the transverse fluctuations associated with this symmetry will
manifest themselves less strongly in the equilibrium system than in
the aging one.

\begin{table}[htb]
\centering
\begin{tabular}{|p{2cm}p{1.5cm}p{1.5cm}p{1.5cm}|}
\hline
System  &   $N_{B_{\vec{r}}}$ & $q_{EA}$ & $\beta$ \\
\hline
Equilibrium & 216 & 0.65 & 0.775 \\
Aging       & 125 & 0.775  & 0.875 \\
\hline
\end{tabular}
\caption{Fitting parameters for coarse graining sizes such that
  $||\delta C^{T}||^2 \approx 0.6 \; ||\delta C||^2$
  in the equilibrium system with $M=94$ and $N_R=10$, and 
  in the aging system with $M=100$ and $N_R=10$.}
\label{tab:parameters}
\end{table}

From now on, we consider in detail one coarse graining volume for each
system. We choose in each case the smallest coarse-graining volume for
which transverse fluctuations capture 60\% of the total variance of
the fluctuations, {\em i.e.\/} $||\delta C^{T}||^2 \ge 0.6 \; ||\delta
C||^2$. This corresponds to $N_{B_{\vec{r}}} = 125$ for the aging
system and $N_{B_{\vec{r}}} = 216$ for the equilibrium
system. Table~\ref{tab:parameters} contains a summary of the fitting
parameters for those coarse graining volumes.

\begin{figure}[h]
  \centerline{ 
    \includegraphics[width=6cm,angle=270]{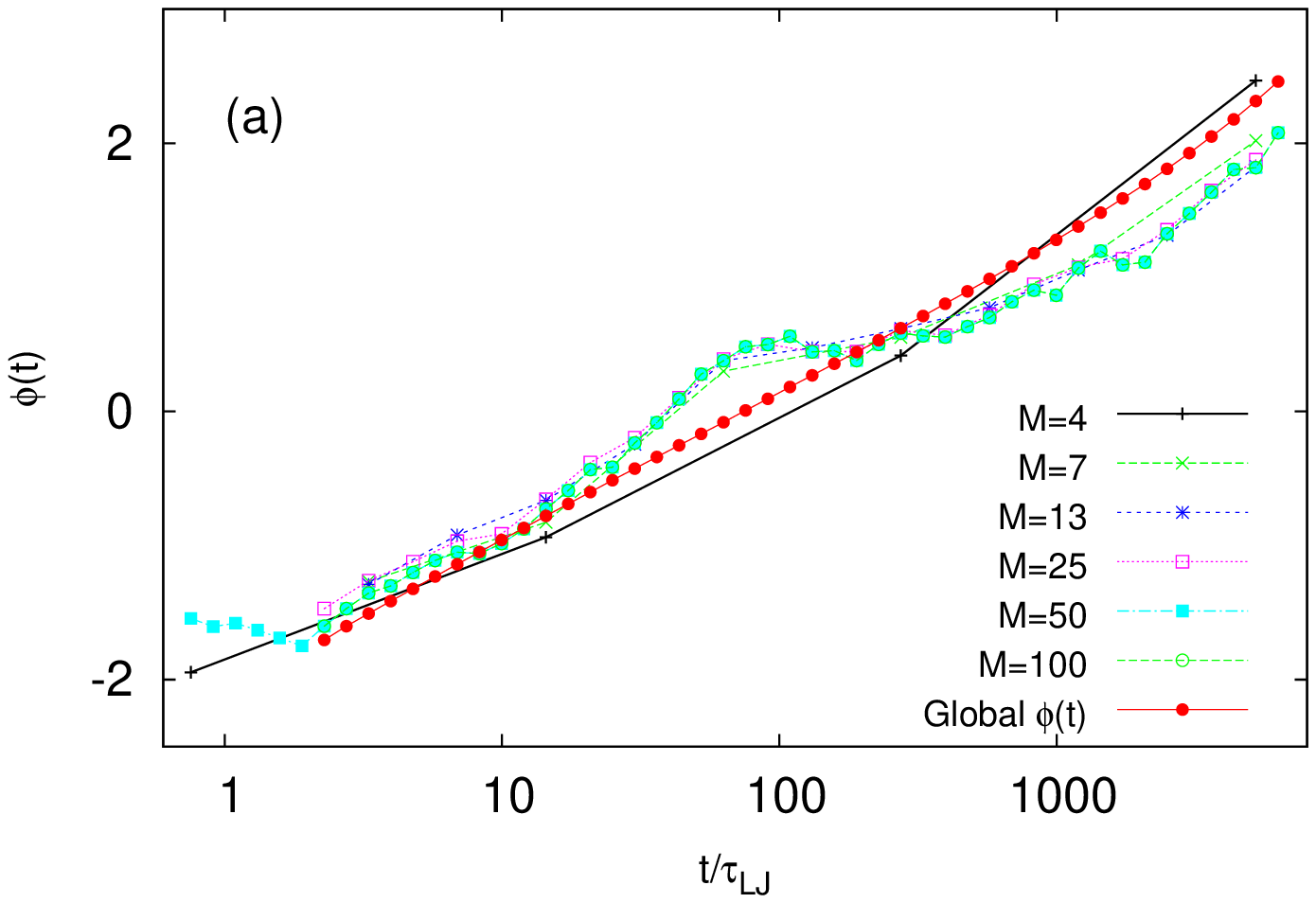} 
    \hspace{0.3cm}
    \includegraphics[width=6cm,angle=270]{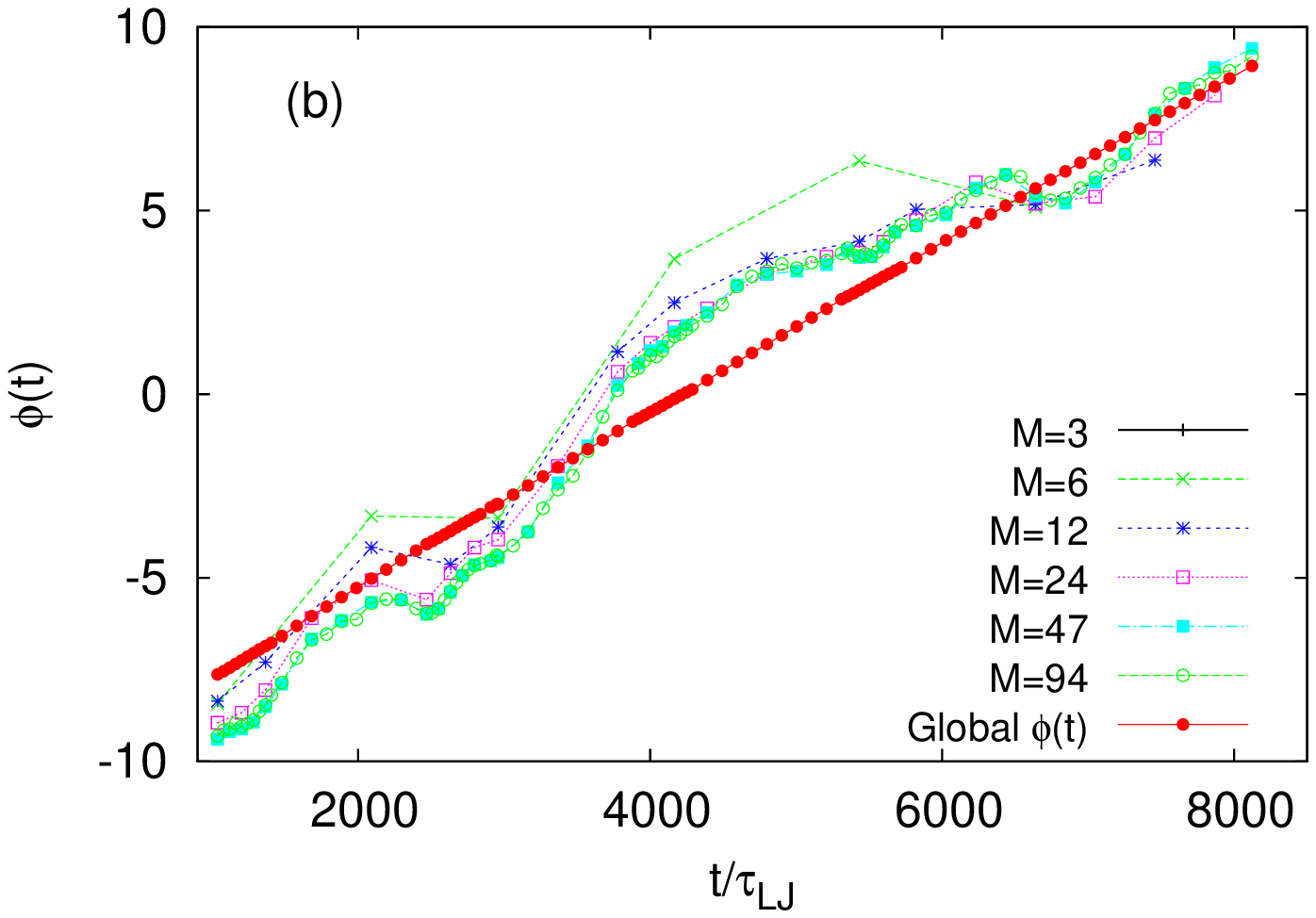}
  }
  \caption{(Color online) Dependence of the function $\phi_{\vec{r}}(t)$ on the
    number of snapshots $M$ used to compute it, for the aging system with
    $N_{B_{\vec{r}}}=125$ (left panel), and for the system 
    in equilibrium with $N_{B_{\vec{r}}}=216$ (right panel).}
  \label{fig:t-slices}
\end{figure}

Since the number of snapshots $M$ used in the analysis cannot be
increased indefinitely, it is important to test that the results are
robust with respect to changes in $M$, and that meaningful results can
be obtained even for relatively small values of $M$. There are two
aspects to this question: one is to show that the phases
$\phi_{\vec{r}}(t)$ are well defined, by showing that their
determination is robust with respect to changes in the number of
snapshots $M$, and the other is to show that the magnitudes of the
transverse, longitudinal and total fluctuations are not singular as a
function of $M$. With regards to the first aspect, we find that as the
number of snapshots is increased, $\phi_{\vec{r}}(t)$ quickly
converges to a relatively smooth function. Fig.~(\ref{fig:t-slices})
shows that $\phi_{\vec{r}}(t)$ changes very little if the number of
snapshots $M$ used to compute it is 7 or larger in the case of the aging
system or if it is 24 or larger in the case of the equilibrium
system. It is noticeable that the minimum number $M_{st}$ of snapshots
needed for a stable determination of $\phi_{\vec{r}}(t)$ is comparable
to the total variation $\Delta \phi$ of the global phase over the
interval considered: in the aging case $M_{st} \approx 7$ and $\Delta
\phi \approx 4$, while in the equilibrium case $M_{st} \approx 24$ and
$\Delta \phi \approx 18$. It is tempting to speculate that at least
one or two configurations per relaxation time are needed in order to
capture a minimum of information about the $\alpha$ relaxation in the
system, and this would lead to the prediction that $M_{st} / \Delta
\phi \sim$ 1--2. However at this point the data we have available are
insufficient to draw any definite conclusions.

\begin{figure}[h]
  \centerline{ 
     \includegraphics[width=6cm,angle=270]{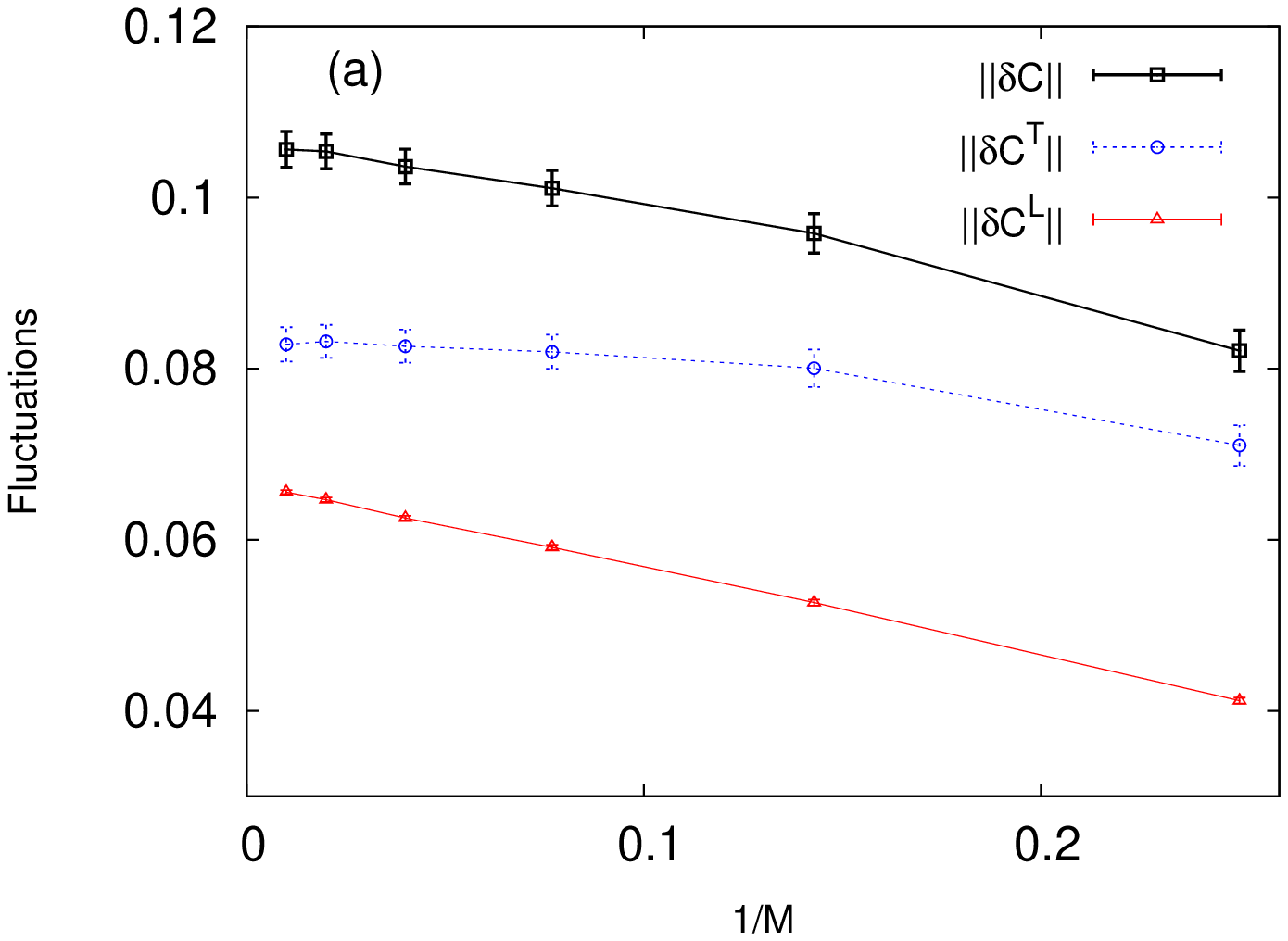} 
      \hspace{0.3cm}
      \includegraphics[width=6cm,angle=270]{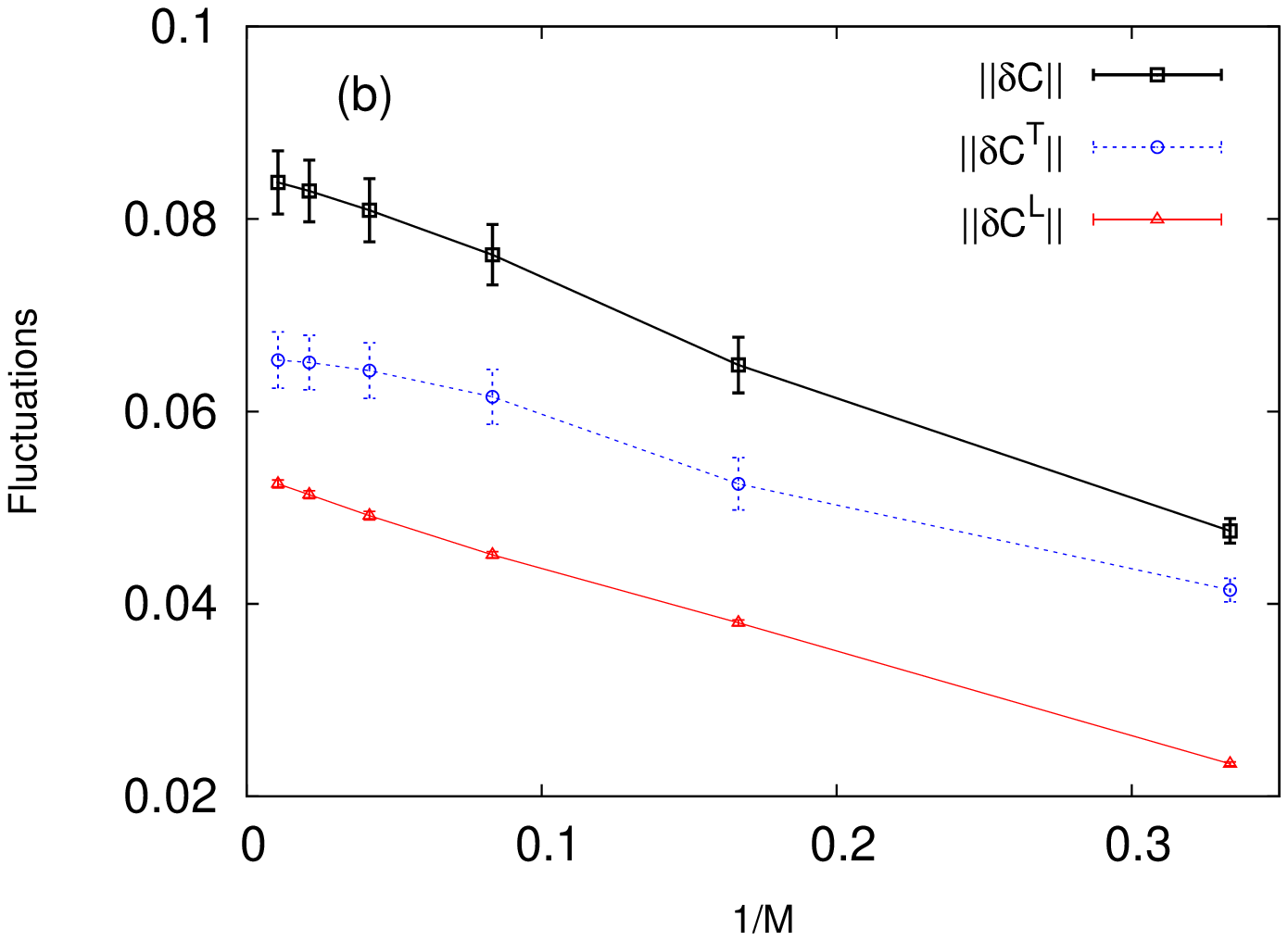}
  }
  \caption{(Color online) Dependence of the magnitude $||\delta C||$ of the total
    fluctuations, the magnitude $||\delta C^{T}||$ of the transverse
    fluctuations, and the magnitude $||\delta C^{L}||$ of the
    longitudinal fluctuations on the number $M$ of snapshots used in
    their evaluation, for the aging system with $N_{B_{\vec{r}}}=125$
    and $N_R=100$ (left panel), and for the system in equilibrium with
    $N_{B_{\vec{r}}}=216$ and $N_R=100$ (right panel). In all cases, the magnitude
    of each kind of fluctuation is a gently increasing function of $M$
    that appears to extrapolates to a well-defined value as $M \to
    \infty$. Error bars are shown, with some of 
    the bars not clearly visible because they are smaller than the 
    symbols.}
  \label{fig:fluct_t-slices}
\end{figure}

We now consider the issue of the smoothness of the dependence of the
magnitudes of the different kinds of fluctuations on the number $M$ of
snapshots. In Fig.~(\ref{fig:fluct_t-slices}) we plot the magnitude
$||\delta C||$ of the total fluctuations, the magnitude $||\delta
C^{T}||$ of the transverse fluctuations, and the magnitude $||\delta
C^{L}||$ of the longitudinal fluctuations as functions of $1/M$. We
find that all three fluctuation magnitudes are gently increasing
functions of $M$, and that each of them appears to approach a constant
as $M \to \infty$. For the specific coarse graining sizes shown in the
figure, the longitudinal fluctuations are always weaker than the
transverse ones, but the opposite case is found for small enough
coarse graining sizes.

\subsection{Local Phases and Local Relaxation Times}
\label{sec:relaxation-times}

Up to this point we have presented evidence in favor of the statement
that the method we are introducing leads to a robust determination of
the transverse and longitudinal components $\delta C^{T}$ and $\delta
C^{L}$ of the fluctuations and of the fluctuating phases
$\phi_{\vec{r}}(t)$. We now use the method to show some examples of
the phases $\phi_{\vec{r}}(t)$ and their time derivatives
$\dot{\phi}_{\vec{r}}(t) \equiv d\phi_{\vec{r}}(t)/dt$. The time
derivatives are particularly interesting, not only because they are
gauge invariant, but also because, according to
Eq.~(\ref{eq:phi-to-tau}), they can in principle be interpreted as
local relaxation rates or inverse local relaxation times
$\tau^{-1}_{\vec{r}}(t)$.

\begin{figure}[h]
  \centerline{ 
     \includegraphics[width=6cm,angle=270]{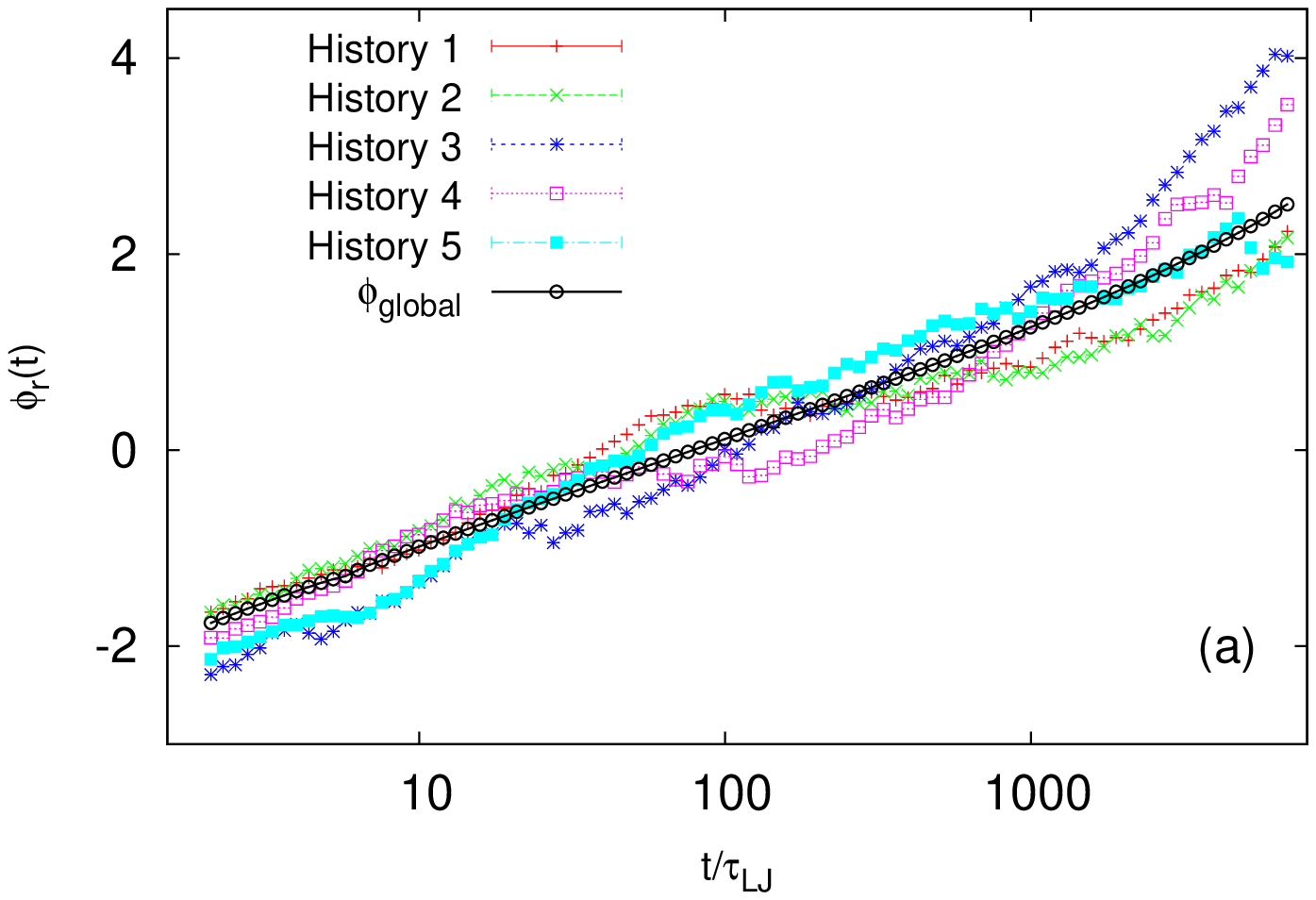} 
      \hspace{0.3cm}
      \includegraphics[width=6cm,angle=270]{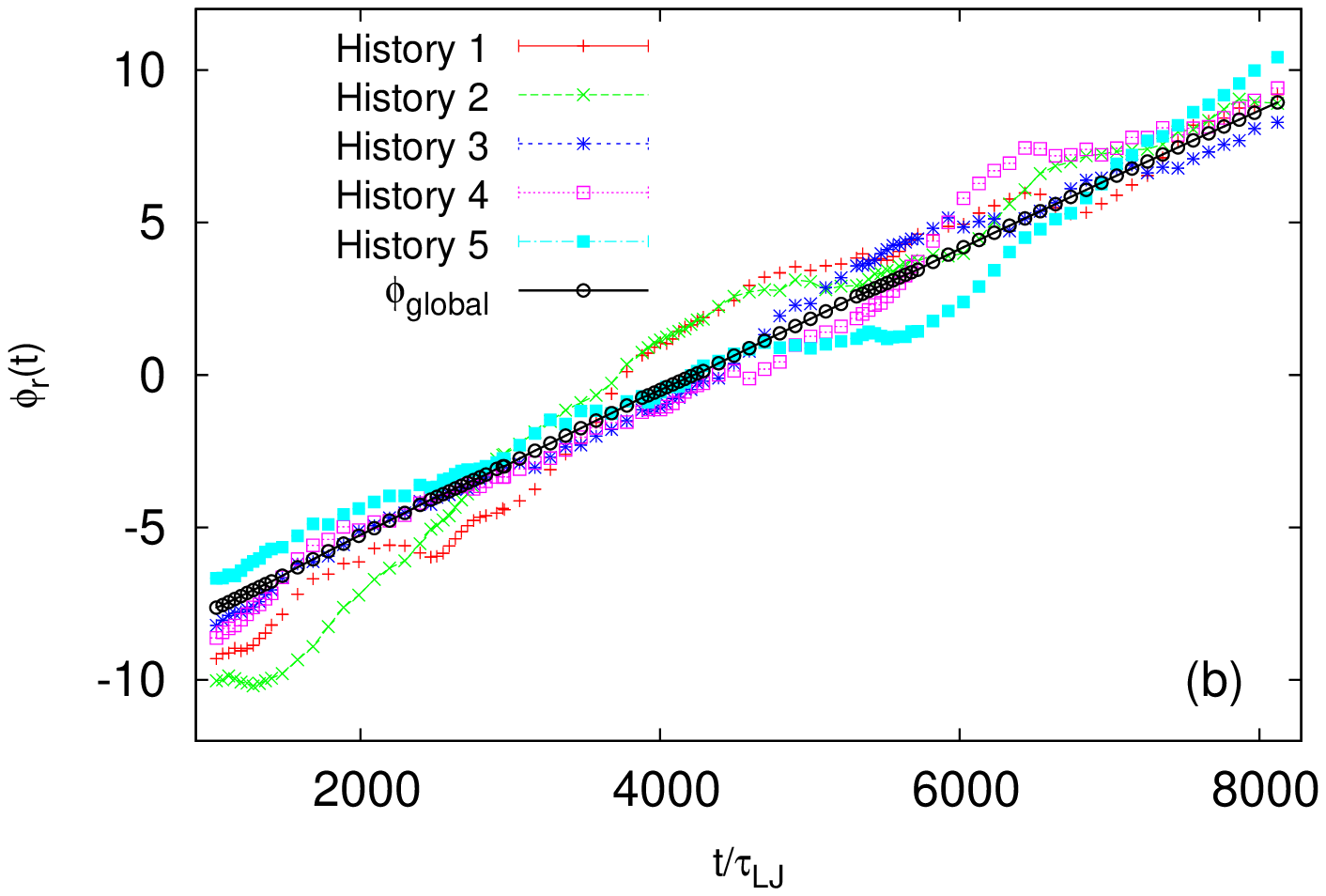}
  }
  \caption{(Color online) Comparison between the local phase
    $\phi_{\vec{r}}(t)$ as a function of time at a fixed point in
    space for different thermal histories and the global phase
    $\phi(t)$. Shown for the aging system, with $N_{B_{\vec{r}}}=125$
    and $M=100$ (left panel), and for the equilibrium system, with
    $N_{B_{\vec{r}}}=216$ and $M=94$ (right panel).  Error bars are
    smaller than the symbols.}
  \label{fig:phase-fluct}
\end{figure}

In Fig.~(\ref{fig:phase-fluct}) we show the time evolution of the local
phases $\phi_{\vec{r}}(t)$ at a fixed point $\vec{r}$ in the sample
for different thermal histories, for both the aging system and the
equilibrium system. For comparison, we also show the evolution of the
global phase $\phi(t)$, and we indeed find that for different thermal
histories the local phase $\phi_{\vec{r}}(t)$ gently fluctuates around
the global value.

\begin{figure}[h]
  \centerline{
     \includegraphics[width=5.7cm,angle=270]{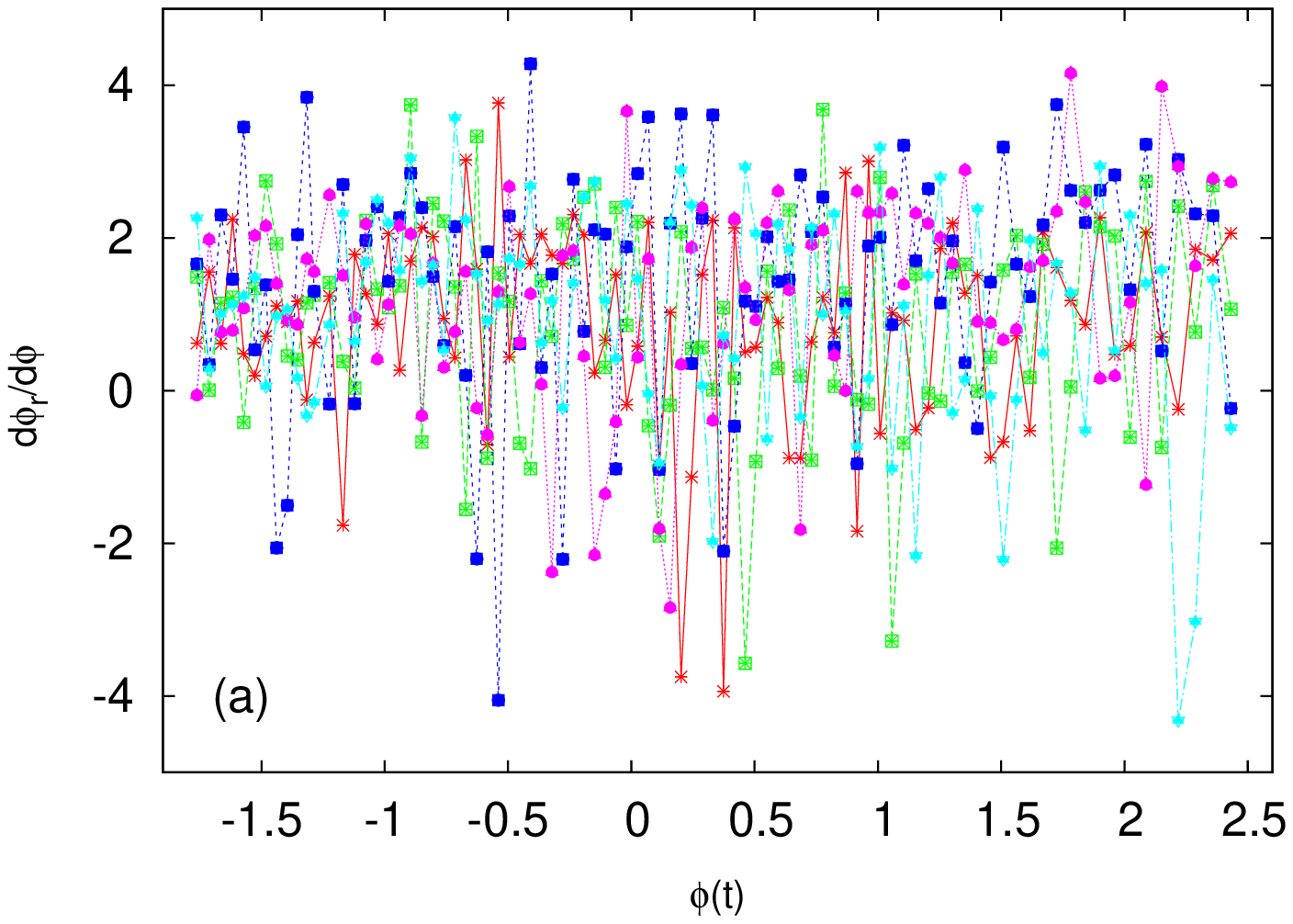}
     \hspace{0.3cm} 
     \includegraphics[width=6cm,angle=270]{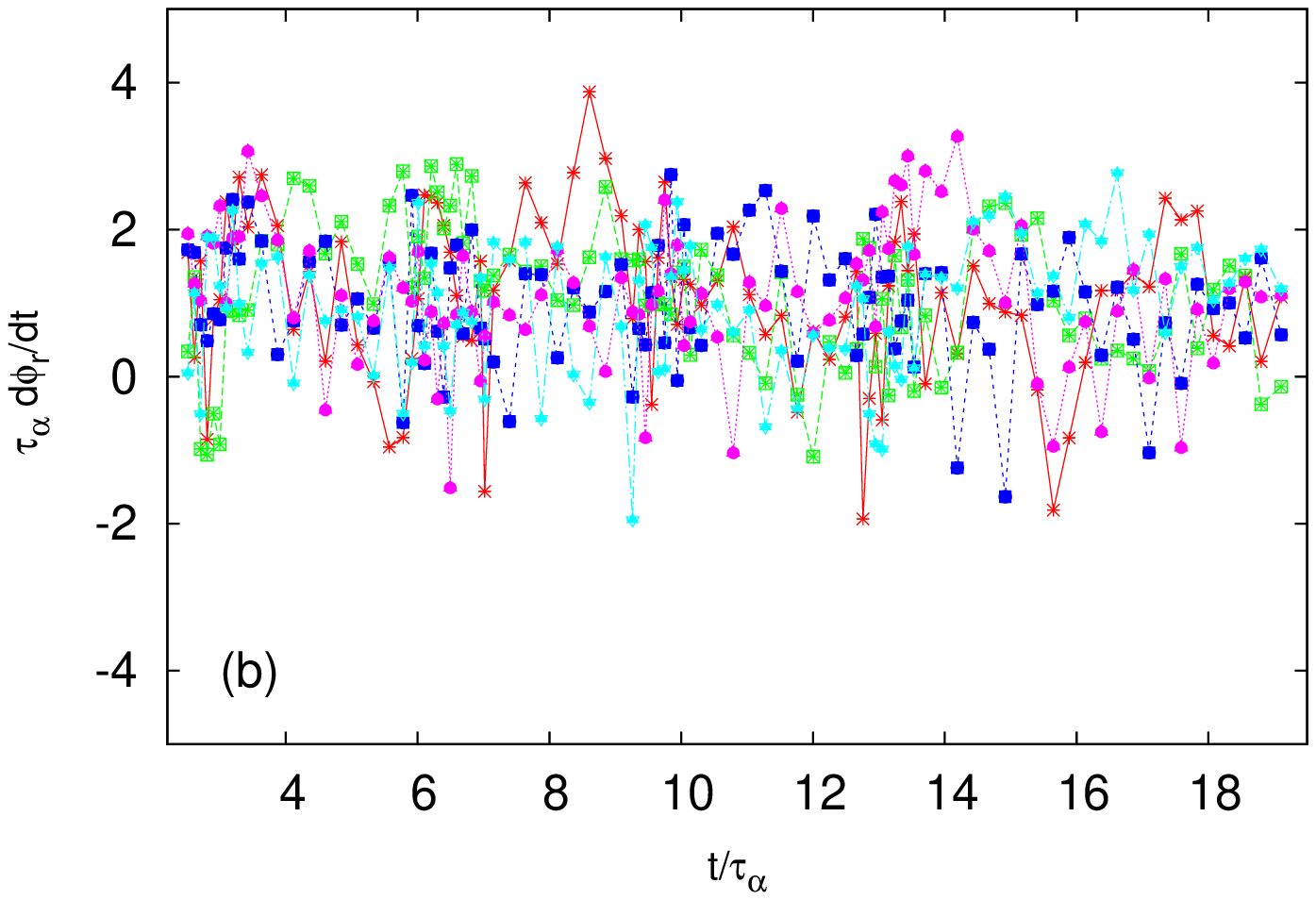}
  }
  \caption{(Color online) Time evolution of the time derivative of the
    local phases $\dot{\phi}_{\vec{r}}$, rescaled so that the time
    derivative of the global phase is unity. In the left panel,
    corresponding to the aging system with $N_{B_{\vec{r}}}=125$ and
    $M=100$, the time derivative of the local phases is rescaled by
    dividing by the global time dependent relaxation rate
    $\frac{d\phi}{dt}$, and it is plotted as a function of the global
    phase $\phi(t)$. In the right panel, corresponding to the
    equilibrium system with $N_{B_{\vec{r}}}=216$ and $M=94$, the
    rescaling of $\dot{\phi}_{\vec{r}}$ is performed approxmately by
    multiplying it with the $\alpha$-relaxation rate $\tau_{\alpha}
    \equiv 1/\overline{d\phi/dt}$, and the rescaled time derivative is
    plotted as a function of the rescaled time $t/\tau_{\alpha}$,
    which is approximately equal to the global phase $\phi(t)$ plus a
    constant. Error bars are smaller than the symbols.  }
  \label{fig:tau-history}
\end{figure}

\begin{figure}[h]
  \centerline{
     \includegraphics[width=10cm,angle=270]{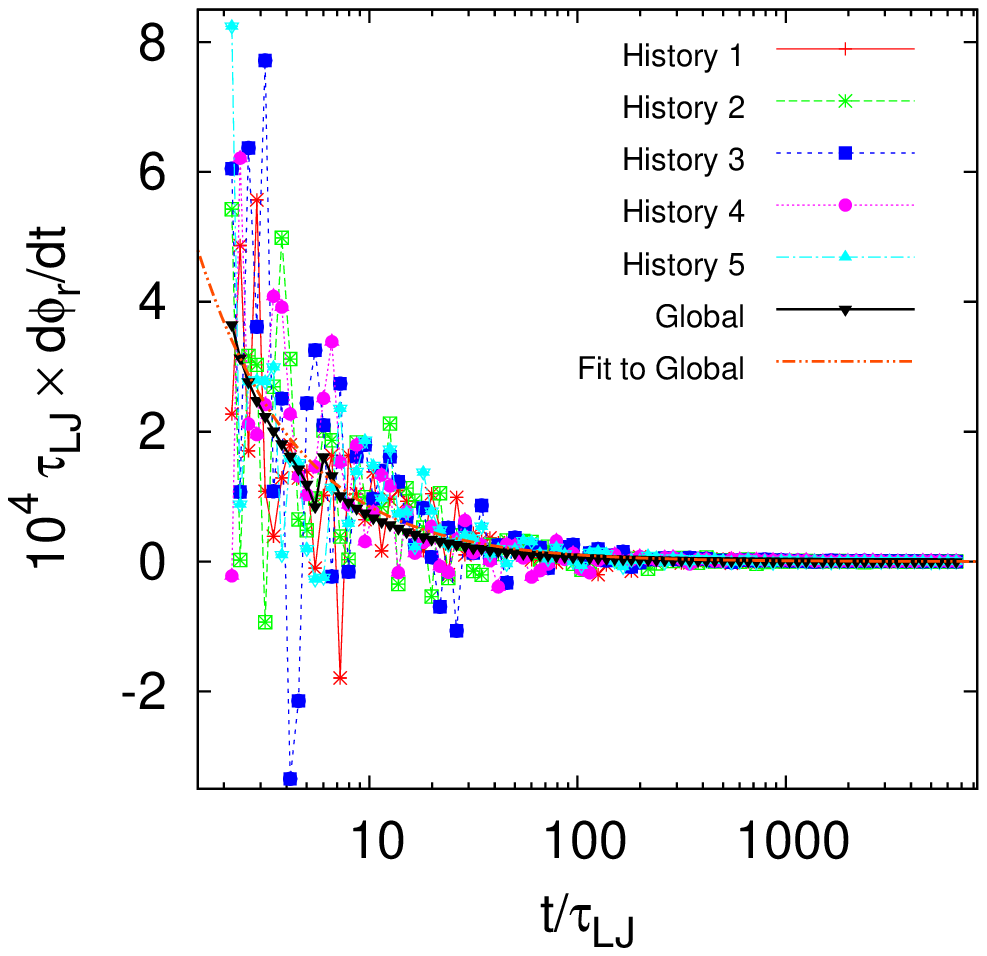}
      }
  \caption{(Color online) Time evolution of the time derivative of the
    local phase, $\dot{\phi}_{\vec{r}}$, for the aging system with
    $N_{B_{\vec{r}}}=125$ and $M=100$. The global
    relaxation rate is also shown, together with a fit of the form
    $d\phi/dt = (t/t_0)^{\alpha}$, where $t_0 \approx 3.4 \times
    10^{-4}$ and $\alpha \approx -0.91$.  Error bars are smaller than
    the symbols.}
  \label{fig:dphi-aging}
\end{figure}

In Figs.~(\ref{fig:tau-history}) and~(\ref{fig:dphi-aging}) we show
the time evolution of the time derivative $\dot{\phi}_{\vec{r}}(t)$ of
the local phase at the same point in space and for the same thermal
histories as in the previous figure. In Fig.~(\ref{fig:tau-history}),
$\dot{\phi}_{\vec{r}}(t)$ is rescaled so that the instantaneous global
relaxation rate is unity, and results are shown both for the aging and
the equilibrium cases. In the case of the equilibrium system, the
rescaling is done approximately, by multiplying the time derivative of
the phase by the $\alpha$-relaxation rate $\tau_{\alpha} \equiv
1 / \overline{d\phi/dt}$, where $\overline{\cdots}$ denotes a time
average, and the fluctuating values of $\dot{\phi}_{\vec{r}}(t)$ are plotted as functions
of the rescaled time $t/\tau_{\alpha}$, which is approximately equal
to the global phase plus a constant. In the case of the aging system,
the rescaling is done by dividing by $d\phi/dt$, and the scaled
values of $\dot{\phi}_{\vec{r}}(t)$ are plotted as functions of the global phase
$\phi(t)$. In both cases the time derivative of the local phase
fluctuates strongly, often becoming more than twice as large as the
global relaxation rate $d\phi/dt$, or even becoming negative. It
should be pointed out that at the times when $\dot{\phi}_{\vec{r}}(t)$
becomes negative, its interpretation as a local relaxation rate
$1/\tau_{\vec{r}}(t)$ becomes problematic. However, from a more
general point of view, it is not surprising that transient
fluctuations in a small region of the sample may give rise to changes
that make the configuration of the region temporarily become closer to
what it was in the past, thus making the two time correlation
$C_{\vec{r}}(t,t')$ increase over time instead of decreasing, with the
effect that, for some time at least, $\dot{\phi}_{\vec{r}}(t)$ becomes
negative. In Fig.~(\ref{fig:dphi-aging}) we plot the time derivative of
the local phase in the aging system, without rescaling it, as a
function of time in Lennard-Jones units. For times longer than $2.5$
Lennard-Jones time units, the global relaxation rate $d\phi/dt$
decreases as a function of time, which is a direct consequence of the
aging in the system, namely the fact that the dynamics becomes slower
as the relaxation progresses. In this figure we can notice again the
strong fluctuations of the time derivative of the local phase with
respect to the global value, but we also notice that both the typical
value and the fluctuations of $\dot{\phi}_{\vec{r}}(t)$ decrease
together with the decrease of $d\phi/dt$. Also shown in the same
figure is a fit of the global relaxation rate by the form $d\phi/dt =
\left(\frac{t}{t_0}\right)^{\alpha}$, where $t_0 \approx 3.4 \times
10^{-4}$ and $\alpha \approx -0.91$, a result that is in agreement
with the results in Refs.~\cite{Avila2011,triangular_long}.
%
%

\section{Summary and Conclusions}
\label{sec:conclusion}

In this work we have shown that a quantitative description of glassy
relaxation in terms of local fluctuating phases and relaxation times
is possible. On one hand, starting from the case of exponential
relaxation we presented a line of phenomenological arguments that
shows how a local phase function $\phi_{\vec{r}}(t)$ can emerge in the
description of the data, and how its time derivative can, under
certain circumstances, be interpreted as a local fluctuating
relaxation rate $1/\tau_{\vec{r}}(t)$. On the other hand, we showed
that in a theoretical framework that postulates the presence of a
broken symmetry under time reparametrizations, it is expected that as
the dynamics becomes more glassy, fluctuations should be dominated by
Goldstone modes - {\em i.e.\/} transverse fluctuations - which are
described naturally in terms of the same local fluctuating phases
$\phi_{\vec{r}}(t)$. Besides establishing a connection between the two
points of view, we have also derived a practical method for extracting
the phases $\phi_{\vec{r}}(t)$ from numerical or experimental data.

We have applied this method to two numerical simulation datasets, one
corresponding to a system undergoing aging at a temperature slightly
below the mode coupling critical temperature $T_c$ and another one
corresponding to a system in equilibrium at a temperature slightly
above $T_c$. We have found that in both cases the results are robust
with respect to changes in the number $M$ of position snapshots used
in the analysis, as long as $M$ is not too small. 

The time reparametrization symmetry framework predicts that for larger
coarse graining regions and lower temperatures the transverse
fluctuations should be dominant, and indeed, this is what we found in
our results. For both systems, the ratio between the magnitude of the
transverse fluctuations and the magnitude of the total fluctuations
grows monotonously with the coarse graining size. For the smallest
coarse graining sizes, longitudinal fluctuations dominate, but the
transverse fluctuations dominate for large coarse graining sizes.
Also, as expected, the range of coarse graining sizes for which
transverse fluctuations dominate is wider in the system that is in
contact with a heat reservoir at a lower temperature than in the
system that is in equilibrium at a higher temperature.

We have shown more detailed results for the fluctuating phase
$\phi_{\vec{r}}(t)$ and its time derivative $\dot{\phi}_{\vec{r}}(t)$,
for a coarse graining size such that the transverse fluctuations
capture slightly more than 60\% of the total fluctuations. For this
coarse graining size, we found that both in the equilibrium system and
in the aging system, the time derivative of the local phase fluctuates rather strongly
over timescales of the order of the instantaneous global relaxation
times or shorter. These fluctuations are strong enough that the time
derivative often takes negative values. The presence of those negative
values, although not surprising, makes an interpretation of
$\dot{\phi}_{\vec{r}}(t)$ as a relaxation rate somewhat
problematic. It is tempting to speculate that this interpretation may
be more cleanly applicable for large enough coarse graining sizes, for
which the total fluctuations are expected to be weak enough that
$\dot{\phi}_{\vec{r}}(t)$ should always remain positive.

Since our method involves fitting the local two-time correlations
$C_{\vec{r}}(t,t')$ by the expression
$g[\phi_{\vec{r}}(t)-\phi_{\vec{r}}(t')]$, one of our results is the
functional form for $g(x)$. In our fits we modeled this function as a
stretched exponential $g(x) = q_{EA}
\exp(-|x|^{\beta})$. Phenomenological arguments have been made that
would indicate that the relaxation should be exponential for small
enough coarse graining regions. If this was the case, we should find
the fitting parameter $\beta$ approaching unity for smaller coarse
graining sizes. Our results in this respect are inconclusive: although
the fitting parameter $\beta$ does approach unity as the coarse
graining size is reduced, and even becomes larger than unity in the
case of the aging system, this happens for a range of coarse graining
sizes in which the fluctuations are mostly longitudinal, or in other
words most of the fluctuations are ``explained'' by the fitting
residuals and not by the fit itself.

One of the ways in which the study of glasses is more difficult than
the study of other condensed matter systems is that in the case of
glasses there are no simple observables that allow probing glassiness
directly. For example, if we want to probe magnetic ordering, a
one-time observable, the magnetization $\langle \vec{S}(\vec{r},t)
\rangle$, is enough to detect the presence of ordering, and magnetic
fluctuations can be probed by the two-point function $\langle
\vec{S}(\vec{r},t) \cdot \vec{S}(\vec{r'},t') \rangle$. For the case
of crystalline ordering, the ordering can be probed by the one-time
Fourier transformed particle density $\rho(\vec{k},t) \equiv N^{-1}
\sum_j \langle \exp[i \vec{k}\cdot\vec{r}_j(t) ] \rangle$, and
correlations can be probed by the dynamic structure factor
$S(\vec{k},t-t') \equiv N^{-1} \sum_{j,j'} \langle \exp[-i
\vec{k}\cdot\vec{r}_{j}(t) ] \exp[i \vec{k}\cdot\vec{r}_{j'}(t') ]
\rangle$. In the case of structural glasses, by contrast, the presence
of a glass state is normally detected by probing the freezing of the
density fluctuations, which already requires a two-time observable
such as the intermediate scattering function $F_s(\vec{q},t,t')$ or
the mean square displacement $\Delta(t,t')$; and the corresponding
fluctuations are quantified by a 4-point correlation function such as
$G_4 (\vec{r},t-t')$ or its Fourier transform $S_4(\vec{q},t,t')$,
which correspond to taking products of two-time observables probing
different spatial points in the system. The description of glassy
dynamics that we have presented here, although it is extracted from
the two-time quantities that probe freezing in the system, has the
advantage that it is expressed in terms of one time quantities, the
local phases $\phi_{\vec{r}}(t)$, or, more properly, their gauge
invariant time derivatives $\dot{\phi}_{\vec{r}}(t)$. This should make
it possible, in principle, to probe correlations in the dynamics by
computing two-point functions, such as $\langle
\dot{\phi}_{\vec{r}}(t) \dot{\phi}_{\vec{r'}}(t') \rangle$, as opposed
to four-point functions. One way of thinking about this is that this
advantage in simplicity is gained by coarse graining both in space and
in time. The coarse graining in space is explicit, and has been
discussed extensively in this work; the coarse graining in time
happens in the sense that the fitting procedure at the heart of the
method focuses on extracting information about the slow part of the
dynamics and ignores the fast part. On one hand, it could be argued
that this coarse graining comes with two kinds of costs: one is the
loss of some information about processes that are ``fast'' in space
and time, and the other is the fact that some extra steps have been
introduced in the analysis of the data. On the other hand, it is
possible that this loss of information and these extra steps in the
analysis may make it easier to observe more cleanly the essential
features of the dynamics.

At this point we have introduced a new general analysis tool to study
glassy dynamics. Since the starting point of this analysis is the
determination of local two-time correlations, it can be applied to any
experiment or numerical simulation data set which provides enough
information to compute those correlations. This includes not only
results from numerical simulations, but also results from
visualization experiments~\cite{Weeks2000, Weeks2002,
  Courtland2003,Keys2007} in which the particle positions are
determined as a function of time. More generally, there are other
kinds of experiments, such as photon correlation
imaging~\cite{Duri2009}, which provide data for local two-time
correlations, and in this case too this type of analysis is
applicable.

The investigation presented here leaves many questions open to be
addressed. Those questions include the determination of the basic
statistical properties of the fluctuating quantity
$\dot{\phi}_{\vec{r}}(t)$, such as its probability distribution and
its correlation functions, the study of how those properties vary with
changes in the control parameter, namely either the temperature or the
volume fraction, and also the study of how they change across
different systems exhibiting glass-like dynamics. In particular, the
correlation function $\langle \dot{\phi}_{\vec{r}}(t)
\dot{\phi}_{\vec{r'}}(t') \rangle$ encodes important information about
dynamical heterogeneity: in the equal-time $t=t'$ case it gives very
direct information about the spatial extent of the heterogeneous
regions, and in the equal-point $\vec{r} = \vec{r'}$ case it gives
information about the persistence in time of those regions. Another
set of questions that would be worth investigating is the relationship
between the present description of the dynamics, and descriptions
based on other probes such as the dynamical susceptibility $\chi_4$ or
the various point-to-set correlation functions. Another set of
questions that could be addressed is a more detailed study of the form
of the function $g(x)$, and in particular the question of the possible
presence of simple exponential relaxation at relatively small coarse
graining scales.  

\begin{acknowledgements}

We thank C.~Chamon, L.~Cugliandolo and M. Kennett for discussions.
This work was supported in part by DOE under grant DE-FG02-06ER46300,
by NSF under grants PHY99-07949 and PHY05-51164, and by Ohio
University. G.~A.~M.~acknowledges the Condensed Matter and Surface
Sciences (CMSS) program for support through a studentship. Numerical
simulations were carried out at the Ohio Supercomputing
center. H.~E.~C. acknowledges the hospitality of the Aspen Center for
Physics and the Kavli Institute for Theoretical Physics, where parts
of this work were performed.

\end{acknowledgements}

\appendix

\section{Derivation of the Method}
\label{app:matrix-method}
Here we present some extra details about the derivation of the method
for extracting local phases from numerical or experimental data, which
we omitted in Sec.~\ref{sec:method}. We start by considering the minimization of
$\bar{\epsilon}(\{\phi_1,\cdots,\phi_M\};\vec{\alpha})$, defined by
Eq.~(\ref{eq:global-epsilon}). By using the definition of $\Phi_{ji}$ in
Eq.~(\ref{eq:Phi-def}), and explicitly writing the terms in the Taylor
expansion $g^{(1)}(\phi_j-\phi_i;\vec{\alpha})$ we rewrite
$\bar{\epsilon}$ to obtain
\begin{eqnarray}
\bar{\epsilon}(\{\phi_1,\cdots,\phi_M\};\vec{\alpha}) 
& = & \eta^{-1}(M) \sum_{1 \le i < j \le M}
\left( g\left(\Phi_{ji};\vec{\alpha}\right) - 
\left\{ g\left(\Phi_{ji};\vec{\alpha}\right)
+ g'\left(\Phi_{ji};\vec{\alpha}\right) \left[ (\phi_j-\phi_i)-
\Phi_{ji} \right] + \cdots  
\right\} \right)^2 \nonumber \\
& = & \frac{1}{2 \eta(M)} \sum_{1 \le i, j\le M}
\left\{ g'\left(\Phi_{ji};\vec{\alpha}\right) \left[ (\phi_j-\phi_i)-
\Phi_{ji} \right] + \cdots  
\right\}^2
\label{eq:global-epsilon-g}  
\end{eqnarray}
where the terms omitted are of quadratic order in
$\left((\phi_i-\phi_j)-\Phi_{ij}\right)$. In the last line we have
allowed both $i>j$ and $i<j$ terms to appear in the sum, and we have
defined $\Phi_{ii} \equiv 0$ so that $i=j$ terms
are identically zero. We now impose the condition that all
derivatives of $\bar{\epsilon}$ with respect to $\phi_k$ be zero, 
\begin{equation}
0 = \frac{\partial\bar{\epsilon}}{\partial\phi_k} = \eta^{-1}(M) \sum_{i=1}^M
g'^2\left(\Phi_{ik};\vec{\alpha}\right) \left[(\phi_i-\phi_k)-
\Phi_{ik}\right], 
\end{equation}
where we have used the fact that $g(-x) = g(x)$ and therefore $g'(-x)
= - g'(x)$. Hence for $1\leq k\leq M$, we have the minimization conditions:
\begin{equation}
\sum_{i=1}^M g'^2\left(\Phi_{ik};\vec{\alpha}\right) \Phi_{ik}
-\sum_{i=1}^M g'^2\left(\Phi_{ik};\vec{\alpha}\right) \phi_i
+\phi_k\sum_{i=1}^M g'^2\left(\Phi_{ik};\vec{\alpha}\right)=0
\end{equation}
We express the minimization conditions in terms of the equivalent matrix equation 
\begin{equation}
\vec{\hat{w}}=\hat{A}\vec{\phi},
\label{eq:global-phase-app}
\end{equation}
with the definition of the M-component vector $\vec{\hat{w}}$
\begin{equation}
\hat{w}_k \equiv \sum_{i=1}^M g'^2\left(\Phi_{ik};\vec{\alpha}\right)
\Phi_{ik}
\end{equation}
and the $M\times M$ matrix $\hat{A}$ 
\begin{equation}
\hat{A}_{ki}\equiv g'^2\left(\Phi_{ik};\vec{\alpha}\right)
-\delta_{ki}\sum_{j=1}^M g'^2\left(\Phi_{jk};\vec{\alpha}\right).
\end{equation}
The system of equations described by Eq.~(\ref{eq:global-phase-app})
is singular, since the nonzero vector $\vec{\phi}_0 \equiv (1,...,1)$
satisfies $\hat{A}\vec{\phi}_0 = \vec{0}$, which indicates that the
rank of the matrix $\hat{A}$ is at most $M-1$. In particular, any
solution $\vec{\phi}$ of Eq.~(\ref{eq:global-phase-app}) can be
modified by adding an arbitrary multiple of $\vec{\phi}_0$ to obtain
another solution. This degeneracy is associated with the gauge
symmetry that the problem has under the shift $\phi_k \to \phi_k +
\rho$, which is a consequence of the fact that in
Eq.~(\ref{eq:local-fit}) the phases appear in a difference. To
eliminate this degeneracy, we impose the additional condition that the
projection of the solution $\vec{\phi}$ on the direction of $\vec{\phi}_0$ should
be zero, thus obtaining the gauge fixing condition of
Eq.~(\ref{eq:g-gauge-cond}). By combining the gauge fixing condition
with Eq.~(\ref{eq:global-phase-app}) above, we obtain a system of
$M+1$ equations with $M$ unknowns:
\begin{equation}
\vec{\hat{w}}=\hat{A}\vec{\phi}\quad\text{ and }\quad 
\vec{\phi}_0\cdot\vec{\phi}=0.
\end{equation}
Since one of the equations 
is a linear combination of the others, the solution is well defined and unique. 
Hence, we omit the first equation, for $k=1$, which is redundant, and solve 
\begin{equation}
\label{eq:global-phase-2-app}
\vec{\bar{w}}=\bar{A}\vec{\phi},
\end{equation}
where $\vec{\bar{w}}$ and $\bar{A}$ are given by 
\begin{eqnarray}
\bar{w}_k & \equiv & \left\{ 
\begin{array}{l@{\quad:\quad}l}
0 & k=1 \\
\hat{w}_k = \sum_{i=1}^M g'^2\left(\Phi_{ik};\vec{\alpha}\right)
\Phi_{ik}
& k \ne 1
\end{array} 
\right.
\nonumber\\
\bar{A}_{ki} & \equiv & \left\{ 
\begin{array}{l@{\quad:\quad}l}
1 & k=1 \\
\hat{A}_{ki} = g'^2\left(\Phi_{ik};\vec{\alpha}\right)
-\delta_{ki}\sum_{j=1}^M
g'^2\left(\Phi_{jk};\vec{\alpha}\right)
& k \ne 1.
\end{array} 
\right.
\end{eqnarray}
which is the same as Eq.~(\ref{eq:bar-w-bar-A}). 

We now turn to the determination of the local phase fluctuations
$\delta \phi_{i \vec{r}}$, which requires minimizing the expression
for $\epsilon$ in Eq.~(\ref{eq:epsilon-C-Taylor}). Here the derivation
is very similar to the previous case. We start by using the definition
of $\delta C_{\vec{r}}(t,t')$ in Eq.~(\ref{eq:dC-def}) to rewrite
$\epsilon$ as
\begin{eqnarray}
\epsilon(\{\delta\phi_{\vec{r}}(t_i)\};
  \vec{\alpha}, \{C_{\vec{r}}(t_j,t_i)\}) =
  \frac{1}{2\eta(M)} \sum_{1\le i, j \le M} \left\{
    \delta C_{\vec{r}}(t_j,t_i)- g'(\phi(t_j)-\phi(t_i);\vec{\alpha})
    \left[\delta\phi_{\vec{r}}(t_j)-\delta\phi_{\vec{r}}(t_i)\right]
    \right\}^2. 
\end{eqnarray}
We then impose the condition that the derivatives with respect to all
of the phase fluctuations $\delta \phi_{k \vec{r}}$ must be zero, 
\begin{equation}
\label{eq:min-condition}
0 = \frac{\partial \epsilon}{\partial \delta \phi_{k \vec{r}}} = - 2
\eta^{-1}(M) \sum_{i=1}^M \left[\delta C_{ki \vec{r}}
- g'\left(\phi_k-\phi_i;\vec{\alpha}\right)
\left(\delta\phi_{k \vec{r}}- \delta\phi_{i \vec{r}}\right)\right]
g'\left(\phi_k-\phi_i;\vec{\alpha}\right). 
\end{equation}
This is equivalent to imposing, for $1\leq k\leq M$, the conditions
\begin{equation}
\label{eq:matrix-eq}
\sum_{i=1}^M g'\left(\phi_k-\phi_i;\vec{\alpha}\right) \delta C_{ki \vec{r}}
= \delta \phi_{k \vec{r}} \sum_{i=1}^M g'^2\left(\phi_k-\phi_i;\vec{\alpha}\right)
- \sum_{i=1}^M g'^2\left(\phi_k-\phi_i;\vec{\alpha}\right)
  \delta\phi_{i \vec{r}}.
\end{equation}
This set of conditions can be written as the matrix equation
\begin{equation}
\tilde{w}_{\vec{r}} = \tilde{A} \delta \phi_{\vec{r}},
\end{equation}
where 
\begin{equation}
\tilde{w}_{k \vec{r}} \equiv \sum_{j=1}^M
g'\left(\phi_k-\phi_j;\vec{\alpha}\right) \delta C_{kj \vec{r}} 
\quad \mbox{and} \quad
\tilde{A}_{ki} \equiv \delta_{ki} \sum_{j=1}^M
g'^2\left(\phi_k-\phi_j;\vec{\alpha}\right)
-g'^2\left(\phi_k-\phi_i;\vec{\alpha}\right).
\end{equation}
From here, the steps are almost identical to those for the
determination of the global phases. Just as in that case, we have a
degeneracy resulting from the shift symmetry under the transformation
$\delta \phi_{k \vec{r}} \to \delta\phi_{k \vec{r}} + \rho_{\vec{r}}$,
where $\rho_{\vec{r}}$ is time independent. We again modify the system
of equations by imposing an additional constraint, given by
Eq.~(\ref{eq:l-gauge-cond}), that removes the degeneracy. Here again
we omit the redundant equation corresponding to $k=1$, and thus
finally recover Eqs.~(\ref{eq:local-phase}) and
(\ref{eq:w-A-local-def}).

\section{Proof of the Orthogonality of $\delta C^T$ and $\delta C^L$}
\label{app:orthogonality}
In this section we prove Eq.~(\ref{eq:T-L-orthogonal}). We start by
applying the definition of the Euclidean scalar product in
Eq.~(\ref{eq:inner-product-def}), 
\begin{equation}
\left( \delta C^{T} | \delta C^{L} \right) = \frac{2}{\omega M(M-1)}
\sum_{1 \le i < j \le M} \sum_{\vec{r}}  \langle 
\delta C^{T}_{\vec{r}}(t_j,t_i) \delta C^{L}_{\vec{r}}(t_j,t_i)\rangle.  
\label{eq:inner-product-CT-CL}
\end{equation}
We now insert the definitions of $\delta C^{T}$ and $\delta C^{L}$, given
by Eqs.~(\ref{eq:dCT-def}) and (\ref{eq:dCL-def}), and obtain
\begin{equation}
\left( \delta C^{T} | \delta C^{L} \right) = \frac{1}{\omega M(M-1)}
\sum_{1 \le j, i \le M} \sum_{\vec{r}} \langle g'(\phi_j - \phi_i)
\left(\delta\phi_{j \vec{r}} - \delta\phi_{i \vec{r}} \right) \left\{
\delta C_{ji \vec{r}} - g'(\phi_j -\phi_i;\vec{\alpha})
\left[\delta\phi_{j \vec{r}} - \delta\phi_{i \vec{r}} \right] \right\}
\rangle,
\label{eq:CT-CL-detail}
\end{equation}
where we have removed the restriction $j>i$. By using the symmetry of
the terms under the exchange of indices $i \leftrightarrow j$, we
rewrite this expression as 
\begin{equation}
\left( \delta C^{T} | \delta C^{L} \right) = \frac{1}{\omega M(M-1)}
\left\langle \sum_{\vec{r}} \sum_{j=1}^M \delta\phi_{j \vec{r}}
\sum_{i=1}^M g'(\phi_j - \phi_i) \left\{ \delta C_{ji \vec{r}} -
g'(\phi_j -\phi_i;\vec{\alpha}) \left[\delta\phi_{j \vec{r}} -
  \delta\phi_{i \vec{r}} \right] \right\} \right\rangle.
\label{eq:CT-CL-rearrange}
\end{equation}
We now recognize that, by Eq.~(\ref{eq:min-condition}), the sum over
the index $i$ is zero, and therefore the whole expression is zero,
which gives us the result that $\left( \delta C^{T} | \delta C^{L}
\right) = 0$.

\end{document}